\documentclass{llncs}
\usepackage{etoolbox}

\usepackage{url}
\usepackage{amssymb,amsthm,latexsym,amsmath,epsfig,amstext}
\usepackage{txfonts}
\usepackage{algorithmic}
\usepackage[linesnumbered,ruled,noend]{algorithm2e}

\usepackage{mathrsfs}

\usepackage{color, colortbl}

\usepackage{todonotes}
\usepackage{multirow}
\usepackage{float,color}
\usepackage{picinpar,color,xcolor,wrapfig}
\newtoggle{full}
\toggletrue{full}

\newcommand{\Nat}{\ensuremath{\mathbb{N}}}

\newcommand{\Int}{\ensuremath{\mathbb{Z}}}

\newcommand{\intnum}{\mathbb{Z}}

\newcommand {\pspace} {\textsc{pspace}}

\newcommand {\expspace} {\textsc{expspace}}
\newcommand {\nexpspace} {\textsc{nexpspace}}

\newcommand{\cut}[1]{}

\mathchardef\mhyphen="2D 

\newcommand{\hide}[1]{}

\newcommand\cB{\mathcal{B}}
\newcommand\cC{\mathcal{C}}

\newcommand\Ll{\mathcal{L}}

\newcommand\cR{\mathcal{R}}

\newcommand\replace{\mathsf{replace}}
\newcommand\replaceall{\mathsf{replaceAll}}

\newcommand\reverse{\mathsf{reverse}}
\newcommand\indexof{\mathsf{indexOf}}
\newcommand\length{\mathsf{length}}
\newcommand\substring{\mathsf{substring}}
\newcommand\charat{\mathsf{charAt}}

\newcommand{\ASSERT}[1]{\textsf{assert}\left(#1\right)}

\newcommand{\concat}{\cdot}

\newcommand{\arity}{r}

\newcommand{\Lang}{\mathscr{L}}
\newcommand{\Tran}{\mathscr{T}}

\newcommand{\NFA}{\mathcal{A}}

\newcommand{\NFT}{\mathcal{T}}

\newcommand{\CEFA}{\mathcal{A}}

\newcommand\bigO{\mathcal{O}}

\newcommand\transducerbench{\textsc{Transducer}}
\newcommand\slogbench{\textsc{SLOG+}}
\newcommand\slogbenchr{\textsc{SLOG+(replace)}}
\newcommand\slogbenchra{\textsc{SLOG+(replaceall)}}
\newcommand\kaluzabench{\textsc{Kaluza}}
\newcommand\pyexbench{\textsc{PyEx}}
\newcommand\pyextdbench{\textsc{PyEx-}td}
\newcommand\pyexztbench{\textsc{PyEx-}z3}
\newcommand\pyexzzbench{\textsc{PyEx-}zz}

\newcommand\slint{${\rm SL}_{\rm int}$}

\newcommand{\regexp} {{\sf RegExp}}

\newcommand{\lasat}{${\rm SAT}_{\rm CEFA}[{\rm LIA}]$}

\newcommand{\uwp}{{\rm uwp}}

\newcommand\urlxsssanitise{{\sf urlXssSanitise}}

\newcommand\cvc{CVC4}
\newcommand\zthree{Z3-str3}
\newcommand\trau{Trau}
\newcommand\trauplus{Trau+}
\newcommand\zthreetrau{Z3-Trau}
\newcommand\ostrich{OSTRICH}
\newcommand\sloth{Sloth}
\newcommand\slent{Slent}
\newcommand\Tool{\text{OSTRICH+}}
\newcommand{\OMIT}[1]{}

\newcommand{\tl}[1]{\color{purple} {TL: #1 :LT} \color{black}}

\title{A Decision Procedure for Path Feasibility of String \\
 Manipulating Programs  with Integer Data Type}

\author{Taolue Chen\inst{1} \and Matthew Hague\inst{2} \and Jinlong He\inst{3,6} \and Denghang Hu\inst{3,6} \and \\ 
	Anthony Widjaja Lin\inst{4} \and Philipp R\"ummer\inst{5} \and Zhilin Wu\inst{3, 7, 8}}

\institute{University of Surrey, UK
	\and Royal Holloway, University of London, UK
	\and State Key Laboratory of Computer Science, \\
	Institute of Software, Chinese Academy of Sciences, China
	\and  Technical University of Kaiserslautern, Germany
    \and Uppsala University, Sweden
    \and University of Chinese Academy of Sciences, China
    \and Shanghai Key Laboratory of Trustworthy Computing, East China Normal University, China
    \and Institute of Intelligent Software, Guangzhou, China
}

\begin{document}
\maketitle

\vspace{-5mm}

\begin{abstract}

\iftoggle{full}{Strings are widely used in programs, especially in web applications.
Integer data type occurs naturally in string-manipulating programs, and is
frequently used to refer to lengths of, or positions in, strings. 
Analysis and testing of string-manipulating programs can be formulated as the 
path feasibility problem:
given a symbolic execution path, does there exist an assignment to the inputs 
that yields a concrete execution that realizes this path?
Such a problem can naturally be reformulated as a string constraint solving
problem.
Although state-of-the-art string constraint solvers usually provide support for both string and integer data types,   
they mainly resort to heuristics without completeness guarantees. \\ 
}{}
In this paper, we propose a decision procedure 
for a class of string-manipulating programs
which includes  not only a wide range of string operations such as concatenation, replaceAll, reverse, and finite transducers, but also those involving the integer data-type such as length, indexof, and substring. To the best of our knowledge, this represents one of the most expressive string constraint languages that is currently known to be decidable.  Our decision procedure is based on a variant of cost register automata. 
We implement the decision procedure, giving rise to a new solver $\Tool$. We evaluate the performance of $\Tool$ on a wide range of existing and new 
benchmarks. The experimental results show that $\Tool$ is the first string 
decision procedure capable of tackling finite transducers and integer constraints, whilst
its overall performance is comparable with the state-of-the-art string constraint solvers.

\end{abstract}

\vspace{-3mm}
\section{Introduction} \label{sec:intro}



String-manipulating programs are notoriously subtle, and their potential bugs 
may bring severe security consequences. A typical example is cross-site scripting
(XSS), which is among the OWASP Top 10 Application Security Risks~\cite{owasp17}.
Integer data type occurs naturally and extensively in string-manipulating programs. 
An effective and increasingly popular method for identifying bugs, including XSS, is symbolic execution~\cite{symbex-survey}.
In a nutshell, this technique analyses static paths
 through the program being considered.
Each of these paths can be viewed as a constraint~$\varphi$ over
appropriate data domains, and symbolic execution tools
demand fast constraint
solvers to check the satisfiability of $\varphi$. Such constraint
solvers need to support all 
data-type operations occurring in
a program.

Typically, mainstream programming languages provide standard string functions such as concatenation, $\replace$, and $\replaceall$. Moreover, Web programming languages usually provide complex string operations (e.g. htmlEscape and trim), which are conveniently modelled as finite transducers, to sanitise malicious user inputs \cite{BEK}. 
Nevertheless, apart from these operations involving only the string data type, functions such as $\length$, $\indexof$, and $\substring$, which can convert strings to integers and vice versa, are also heavily used in practice; for instance, it was reported~\cite{Berkeley-JavaScript} that $\length$, $\indexof$, $\substring$, and variants thereof, comprise over 80\% of string function occurrences in 18 popular JavaScript applications, notably outnumbering concatenation. The introduction of integers exacerbates the intricacy of string-manipulating programs, and poses new theoretical and practical challenges in solver development.

When combining strings and integers, decidability can easily be lost; 
 for instance, the string theory with concatenation and letter counting
functions is undecidable~\cite{buchi,Manea-RP}.
\OMIT{
Although a great deal of research has shown that it is viable to reason about rather complex string operations without breaking decidability by 
restricting the form of constraints (e.g.,~\cite{CCH+18,CHL+19}; cf.\ related work for a brief survey), adding length constraints would immediately lead to undecidability~\cite{CCH+18}. 
}
Remarkably, it is still a major open problem whether the string theory with concatenation (arguably the simplest string operation) and length function 
(arguably the most common string-number function) is 
decidable~\cite{Vijay-length,LinM18}. 
One promising approach to retain decidability is to enforce a syntactic
restriction to the constraints. In the  literature, these restriction  include solved forms
\cite{Vijay-length}, acyclicity \cite{BFL13,Abdulla14,AbdullaA+19}, and 
straight-line fragment (aka programs in single static assignment form) \cite{LB16,CCH+18,CHL+19,HJLRV18}. On the one hand,
such a restriction has led to decidability of string constraint solving with 
complex string
operations (not only concatenation, but also finite transducers) and integer
operations (letter-counting, $\length$, $\indexof$, etc.); see, e.g., 
\cite{LB16}. On the other hand, there is a lot of evidence (e.g. from 
benchmark) that 
many practical string constraints do satisfy 
such syntactic restrictions.
%
%
\OMIT{
Although there have been some recent results on decision procedures for string constraints involving the integer data-type, for instance~\cite{Vijay-length,LeH18,LinM18,LB16}, the theories considered in those approaches are usually quite restricted and do not reflect the constraints that occur in applications well. Overall, much more research is needed to
better understand the decidability of theories combining strings and integers.
}

Approaches to building practical string solvers could essentially be classified
into two categories. Firstly, one could support as many constraints as possible, but 
primarily resort to heuristics, 
offering no completeness/termination guarantee. This is a realistic approach 
since, as
mentioned above, the problem involving both string and integer data types is in 
general undecidable. Many solvers
belong to this category, e.g., 
{\cvc}~\cite{cvc4}, Z3 \cite{BTV09,Z3}, {\zthree}~\cite{Z3-str3}, S3(P) \cite{S3,TCJ16},
Trau~\cite{Abdulla17} (or its variants {\trauplus}~\cite{AbdullaA+19} and {\zthreetrau}~\cite{Z3-trau}), ABC \cite{ABC}, and \slent~\cite{WC+18}.
%
%
\OMIT{
Such solvers
often show excellent performance in applications and on existing
benchmark suites, but 
}
Completeness guarantees are, however, valuable since 
the 
performance of heuristics can be difficult to predict.
The second approach is to develop solvers for decidable fragments
supporting both strings and integers (e.g. 
\cite{Vijay-length,BFL13,Abdulla14,AbdullaA+19,LB16,CCH+18,CHL+19,HJLRV18}). 
Solvers in this category include Norn \cite{Abdulla14}, SLOTH
\cite{HJLRV18}, and OSTRICH \cite{CHL+19}. 
The fragment \emph{without} complex string operations (e.g. $\replaceall$ and
finite transducers, but $\length$) can be handled quite well by Norn. The fragment \emph{without} length
constraints (but $\replaceall$ and finite transducers)
can be handled effectively by OSTRICH and SLOTH. 
Moreover, most existing solvers that belong to the first category do not support
complex string operations like $\replaceall$ and finite transducers as well.
This motivates the following problem: 
\emph{provide a decision procedure that supports both string and integer data type, with completeness guarantee and meanwhile admitting efficient implementation}.

We argue that this problem is highly challenging.
A deeper examination of the algorithms used by OSTRICH and SLOTH reveals that,
unlike the case for Norn, it would \emph{not} be straightforward to extend OSTRICH and 
SLOTH with integer constraints. First and foremost, the complexity of the
fragment used by Norn (i.e. without transducers and $\replaceall$) is solvable
in exponential time, even in the presence of integer constraints. This is not
the case for the straight-line fragments with transducers/$\replaceall$, which 
require at least double exponential time (regardless of the integer constraints).
This unfortunately manifests itself in the size of symbolic representations of 
the solutions. SLOTH \cite{HJLRV18} computes 
a representation of all
solutions ``eagerly'' as (alternating) finite transducers. Dealing with integer data type requires  
to compute the Parikh images of these transducers \cite{LB16}, which would 
result in a
quantifier-free linear integer arithmetic formula (LIA for short) of double exponential size, thus giving us a triple exponential time algorithm, since LIA formulas are solved in exponential time  (see e.g. \cite{VSS05}).
Lin and Barcelo \cite{LB16} provided a double exponential upper bound in the 
length of the strings in the solution, and showed that the double
exponential time theoretical complexity could be retained. This, however, does
not result in a practical algorithm since it requires all strings of double
exponential size to be enumerated. OSTRICH \cite{CHL+19} adopted a ``lazy'' approach and computed the pre-images of regular languages step by step,
which is more scalable than the ``eager'' approach adopted by SLOTH and results in a highly competitive solver.
It uses \emph{recognisable relations} (a finite union of products of regular languages)
as symbolic representations. Nevertheless, extending this approach to integer
constraints is not obvious since integer constraints break the independence
between different string variables in the recognisable relations.

\OMIT{
There is still a gap between the string operations supported by
solvers, and the operations occurring in applications; for instance,
operations like \textsf{replaceAll} and escape/unescape-transformations
occur extremely frequently in programs, but are currently not handled well
by many solvers. 
The \emph{grand challenge} in string constraint solving is to
identify fragments and develop algorithms, that cover the constraints
occurring in \emph{applications}, while preserving \emph{decidability} and
admitting \emph{efficient implementations.}
}

\noindent\emph{Contribution.} We provide a decision procedure for an expressive class of string constraints involving the integer data type, which includes not only concatenation, $\replace$/$\replaceall$, $\reverse$, finite transducers, and regular constraints, but also $\length$, $\indexof$ and $\substring$. The decision procedure utilizes a variant of cost-register automata introduced by Alur et al. \cite{RLJ+13}, which are called \emph{cost-enriched finite automata} (CEFA) for convenience. 
Intuitively, each CEFA records the connection between a string variable and its associated integer variables.
With CEFAs, the concept of recognisable relations is then naturally extended to accommodate integers. The integer constraints, however, are detached from CEFAs rather than being part of CEFAs. This allows to preserve the independence of string variables in the recognisable relation. 
The crux of the decision procedure is to compute the backward images of CEFAs under string functions, where each cost register (integer variable) might be split into several ones, thus extending but still in the same flavour as OSTRICH for string constraints \emph{without} the integer data type \cite{CHL+19}. 
Such an approach 
is able to treat 
a wide range of string functions in a generic, and yet simple, way. 
To the best of our knowledge, the class of string constraints considered in this paper is currently one of the most expressive string theories involving the integer data type known to enjoy a decision procedure. 
%
\OMIT{
Moreover, 
our decision procedure takes a ``local'' approach for step-by-step backward 
computation of pre-images of CEFAs, which extends the framework of {\ostrich}
solver \cite{CHL+19}.

compared to the ``global'' approach in~\cite{LB16} which is based on enumerating
all strings up to a double exponential length (which does not lead to an
efficient implementation), 
}

We implement the decision procedure based on  the recent {\ostrich} solver \cite{CHL+19},  resulting in {\ostrich}+.  We perform experiments on a wide range of benchmark suites, including those where both $\replace$/$\replaceall$/finite transducers and $\length$/$\indexof$/$\substring$ occur, as well as the well-known benchmarks {\kaluzabench} and {\pyexbench}.
The results show that  
1) {\ostrich}+ so far is the only string constraint solver capable of dealing with finite transducers and integer constraints, and 2) its overall performance is comparable with the best state-of-the-art string constraint solvers (e.g. {\cvc} and {\zthreetrau}) which are short of completeness guarantees. 

\hide{
\smallskip

%
%
%
%
%
 
\noindent\emph{Related work.}
Many theoretical results and practical solvers have been mentioned before. We shall discuss some further theoretical results on string constraints involving the integer data type.
%
%
%
%
\hide{
The research on the theory of strings with concatenation 
was motivated by 
the question of solving satisfiability of word equations (i.e., equations containing concatenation of string constants and variables). Makanin showed decidability \cite{Makanin}, but the upper bound was only improved to PSPACE about 25 years later \cite{P04},    
a simpler algorithm of which was given recently \cite{Jez17}. 
The best lower bound for this problem is still NP, and closing this complexity gap is a long-standing open problem. 
The PSPACE upper bound stands 
in the presence of regular constraints (e.g., see \cite{Schulz}). 
%
In practice, it is common for a string-manipulating program to contain multiple operations (e.g., $\replaceall$) that may be well beyond concatenation, which has motivated research on more expressive string constraint languages.   
Unfortunately, it takes very little for 
them to become undecidable \cite{LB16}. For instance, when finite transducers are allowed, checking the satisfiability of a simple formula of the form $\NFT(x, x)$ for a given finite transducer $\NFT$, can easily encode the Post Correspondence Problem, and therefore is undecidable.
%
%

Hence,  recent research endeavors 
to recover decidability of some string constraint languages with multiple string operations, while retaining applicability for constraints that
arise in practical symbolic execution applications. This is mainly done by imposing syntactic restrictions along the way in which string equality can be used in the constraints. This includes acyclicity \cite{Abdulla14,BFL13}, solved form \cite{Vijay-length}, and straight-line \cite{LB16,HJLRV18,CCH+18}.
Particularly related to the present work, the straight-line logic \cite{LB16}  unified the earlier logics by allowing concatenation, regular constraints, rational
transductions. 
It was further extended 
by a more general form of replaceAll functions with the replacement string provided as a
\emph{variable} \cite{CCH+18} where it is shown that the new straight-line logic 
is decidable.  Furthermore, two general semantic conditions are identified  \cite{CHL+19}  which together entail the decidability of path
feasibility. In particular, they are satisfied by a multitude of string operations
including concatenation, one-way and two-way finite-state transducers, (general) replaceAll functions
string-reverse functions, regular-expression matching, etc. 
}
%
Lin and Majumdar, as well as Loc and He, investigated decidable fragments of 
quadratic word equations, linear arithmetic with the length function, and regular constraints~\cite{LinM18,LeH18}. Chen \emph{et al.}\ showed that, although the straight-line string logic with the general form of $\replaceall$ (where the replacement parameter can be string variables) is decidable, it would become undecidable when extended with any of the length, $\indexof$, $\charat$ functions~\cite{CCH+18}. This result implies that the $\replaceall$ function considered in this paper must not be in the general form (i.e., string variables in the replacement string must be dismissed) to preserve the decidability. Chain-free string constraints are another decidable class of string constraints involving integer data type, proposed in \cite{AbdullaA+19}. They are incomparable to our class of string constraints, as they do not support $\indexof$ or $\substring$, but go  beyond the straight-line fragment.

\hide{
\smallskip
\noindent
\textbf{Practical solvers.}
Many practical (often incomplete) string constraint solvers have been developed, where various heuristics are used to deal with strings and integers. 

The solvers Kaluza~\cite{Berkeley-JavaScript} and Hampi~\cite{HAMPI}  bound the lengths of strings and reduce the problem to the satisfiability of boolean formulas or formulae in the theory of bit vectors. Sushi~\cite{sushi} and Stranger~\cite{YABI14,Stranger} use finite automata to over approximate sets of values of string variables. Moreover, SLOG and {\slent}  use symbolic approaches to enhance the scalability of the string constraint solving process \cite{fang-yu-circuits,WC+18}. In contrast, Trau as well as its variants {\zthreetrau} and {\trauplus} utilise flat automata to under-approximate sets of values of string variables \cite{Abdulla17,AbdullaACDHRR18-trau,AbdullaA+19,Z3-trau}.
Instead of using automata, Z3/Z3-str/Z3-str2/Z3-str3~\cite{BTV09,Z3,Z3-str,Z3-str2,Z3-str3} and {\cvc}~\cite{cvc4,ReynoldsWBBLT17} apply rewrite or algebraic rules to solve the string constraints.
Furthermore, S3 and S3P rely on recursive definitions to represent string functions and constraints, which are then unfolded during the constraint solving process \cite{S3,TCJ16}.
Finally, ACO-Solver combined the ant colony optimisation meta-heuristic with automata-based string constraint solvers Sushi~\cite{sushi},  in order to support reasoning about complex string operations related to XSS vulnerabilities \cite{ThomeSBB17}.
}

\hide{
A large body of these tools provide support for complex string operations, but only offer relatively limited support for integer constraints (e.g., only the length function or simple variants thereof). Moreover, most of them rely heavily on heuristics which are short of theoretical guarantees. Some practical heuristics include bounding string lengths (e.g., \cite{HAMPI,Berkeley-JavaScript,BTV09}),  induction, overapproximations \cite{Stranger,YABI14}, interpolation \cite{Abdulla14}, and flat automata \cite{Abdulla17}, context-dependent simplification \cite{ReynoldsWBBLT17}, to name a few. 
Focusing on semi-algorithms also allows highly expressive (but undecidable) 
string  constraint languages, e.g., recursively defined functions \cite{S3,TCJ16}; sometimes hybrid constraint solving procedures based on, for instance, 
ant colony optimization meta-heuristic \cite{ThomeSBB17} is utilized.  
Recently, the study of decidability of string constraints has also resulted in automata-theoretic algorithms that are amenable to implementation. For instance, the tool Norm supports solving  acyclic logic with concatenation, regular constraints, and length constraints \cite{Abdulla14}; 
{\sloth} \cite{HJLRV18} and {\ostrich} \cite{CHL+19} support straight-line logic with finite transducers (or
replaceall), concatenation, and regular constraints, but currently no support for even the length constraints. 
The current work builds on the solver {\ostrich}, but significantly extends its power of handling integer related constraints 
implementing  solving techniques for the most expressive \emph{decidable} string constraints involving both strings and integers. 
}

}

The rest of the paper is structured as follows: Section~\ref{sec:prel} introduces the preliminaries. Section~\ref{sec:logic} defines the class of string-manipulating programs with integer data type. Section~\ref{sec:dec} presents the decision procedure. Section~\ref{sec:eva} presents the benchmarks and experiments for the evaluation. The paper is concluded in Section~\ref{sec:conc}. 
\iftoggle{full}{Missing proofs, implementation details and further examples can be found in the appendix. }{Missing proofs, implementation details and further examples can be found in the full version \cite{atva20-full}.}


\vspace{-1mm}
\section{Preliminaries}\label{sec:prel}


We write $\Nat$ and $\Int$ for the sets of natural and integer numbers, respectively. For $n \in \Nat$ with $n \ge 1$, $[n]$ denotes $\{1, \ldots, n\}$; for $m,n \in \Nat$ with $m \le n$,  $[m, n]$ denotes $\{ i \in \Nat \mid m \le i \le n \}$. Throughout the paper, $\Sigma$ is a finite alphabet, ranged over by $a,b,\ldots$.  

\smallskip
\noindent \emph{Strings, languages, and transductions.}
A string over $\Sigma$ is a (possibly empty) sequence of elements from $\Sigma$,
denoted by $u, v, w, \ldots$. An empty string is denoted by $\varepsilon$.  We write $\Sigma^*$ (resp., $\Sigma^+$) for the set of all (resp. nonempty) strings over $\Sigma$.
For a string $u$, we use $|u|$ to denote the number of letters in $u$. In particular, $|\varepsilon|=0$. Moreover, for $a \in \Sigma$, let $|u|_a$ denote the number of occurrences of $a$ in $u$. 
Assume $u=a_0\cdots a_{n-1}$ is nonempty and $i<j \in [0,n-1]$. 
We let $u[i]$ denote $a_i$ and $u[i,j]$ for the substring 
$a_i\cdots a_j$. 

Let $u, v$ be two strings. We use $u \cdot v$ to denote the \emph{concatenation} of $u$ and $v$.
The string $u$ is said to be a \emph{prefix} of $v$ if $v = u \cdot v'$ for some string $v'$.
In addition, if $u \neq v$, then $u$ is said to be a \emph{strict} prefix of $v$. 
If 
$v = u \cdot v'$ for some string $v'$, then 
we use $u^{-1} v$ to denote $v'$. In particular, $\varepsilon^{-1} v = v$.
If $u=a_0 \cdots a_{n-1}$ is nonempty, then we use $u^{(r)}$ to denote the \emph{reverse} of $u$, that is, $u^{(r)}= a_{n-1 }\cdots a_0$. 


A \emph{transduction} over $\Sigma$ is a binary relation over $\Sigma^*$, namely, a subset of $\Sigma^* \times \Sigma^*$. We will use $T_1, T_2,\ldots$ to denote transductions. For two transductions $T_1$ and $T_2$, we will use $T_1 \cdot T_2$ to denote the \emph{composition} of $T_1$ and $T_2$, namely, $T_1 \cdot T_2 = \{(u, w) \in \Sigma^* \times \Sigma^* \mid \emph{there exists } v \in \Sigma^* \mbox{ s.t. } (u,v) \in T_1 \mbox{ and } (v,w) \in T_2\}$.

\smallskip
\noindent \emph{Recognisable relations.} 
We assume familiarity with standard regular language. Recall that a regular language $L$ can be represented by a regular expression $e\in \regexp$ whereby we usually write $ L=\Ll(e)$. 

Intuitively, a \emph{recognisable relation} is simply a finite union of Cartesian products of regular languages. Formally, an $\arity$-ary relation $R\subseteq \Sigma^*\times \cdots\times \Sigma^*$ is \emph{recognisable} if $R=\bigcup_{i=1}^n L^{(i)}_1\times \cdots\times L^{(i)}_\arity$ where $L^{(i)}_j$ is regular for each $j\in [\arity]$. A \emph{representation} of a recognisable relation $R=\bigcup_{i=1}^n L^{(i)}_1\times \cdots\times L^{(i)}_\arity$ is $(\NFA^{(i)}_1, \ldots, \NFA^{(i)}_\arity)_{1 \le i \le n}$ such that each $\NFA^{(i)}_j$ is an NFA with $\Lang(\NFA^{(i)}_j)=L^{(i)}_j$. The tuples $(\NFA^{(i)}_1, \ldots, \NFA^{(i)}_\arity)$ are called the \emph{disjuncts} of the representation and the NFAs $\NFA^{(i)}_j$ are called the \emph{atoms} of the representation.




\smallskip
\noindent \emph{Automata models.} 
A \emph{(nondeterministic) finite automaton} (NFA)  is a tuple $\NFA=(Q, \Sigma, \delta, I, F)$, where $Q$ is a finite set of states, $\Sigma$ is a finite alphabet, $\delta \subseteq Q \times \Sigma \times Q$ is the transition relation, $I,F \subseteq Q$ are the set of initial and final states respectively. For readability, we write a transition $(q, a, q') \in \delta$ as $q \xrightarrow[\delta]{a} q'$ (or simply $q \xrightarrow{a} q'$). 
The \emph{size} of an NFA $\NFA$, denoted by $|\NFA|$, is defined as the number of transitions of $\NFA$.
A \emph{run} of $\NFA$ on a string $w = a_1 \cdots a_n$ is a sequence of transitions $q_0 \xrightarrow{a_1} q_1 \cdots q_{n-1} \xrightarrow{a_n} q_n$ with $q_0 \in I$. The run is \emph{accepting} if $q_n \in F$.
A string $w$ is accepted by an NFA $\NFA$ if there is an accepting run of $\NFA$ on $w$. In particular, the empty string $\varepsilon$ is accepted by $\NFA$ if $I \cap F \neq \emptyset$. The language of $\NFA$, denoted by $\Lang(\NFA)$, is the set of strings accepted by $\NFA$. 
An NFA $\NFA$ is said to be \emph{deterministic} if $I$ is a singleton and, for every $q \in Q$ and $a \in \Sigma$, there is at most one state $q' \in Q$ such that $(q, a, q') \in \delta$.
It is well-known that finite automata capture regular languages precisely.

A \emph{nondeterministic finite transducer} (NFT) $\NFT$ is an extension of NFA with outputs. Formally, an NFT $\NFT$ is a tuple $(Q, \Sigma, \delta, I, F)$, where $Q, \Sigma, I, F$ are as in NFA and the transition relation $\delta$ is a finite subset of $Q \times \Sigma \times Q \times \Sigma^*$. Similarly to NFA, for readability, we write a transition $(q, a, q', u) \in \delta$ as $q \xrightarrow[\delta]{a, u} q'$ or $q \xrightarrow{a, u} q'$. The \emph{size} of an NFT $\NFT$, denoted by $|\NFT|$, is defined as the sum of the sizes of the transitions of $\NFT$, where the size of a transition $q \xrightarrow{a, u} q'$ is defined as $|u|+3$.
A run of $\NFT$ over a string $w=a_1 \cdots a_n$ is a sequence of transitions $q_0 \xrightarrow{a_1, u_1} q_1 \cdots q_{n-1} \xrightarrow{a_n, u_n} q_n$ with $q_0 \in I$. The run is accepting if $q_n \in F$. The string $u_1 \cdots u_n$ is called the output of the run. The transduction defined by $\NFT$, denoted by $\Tran(\NFT)$, is the set of string pairs $(w, u)$ such that there is an accepting run of $T$ on $w$, with the output $u$. An NFT $\NFT$ is said to be \emph{deterministic} if $I$ is a singleton, and, for every $q \in Q$ and $a \in \Sigma$  there is at most one pair $(q', u) \in Q \times \Sigma^*$ such that $(q, a, q', u) \in \delta$.
In this paper, we are primarily interested in \emph{functional} finite transducers (FFT), i.e., finite transducers that define functions instead of relations. (For instance, deterministic finite transducers are always functional.)





We will also use standard  quantifier-free/existential \emph{linear integer arithmetic} (LIA) formulae, which are typically ranged over by $\phi, \varphi$, etc. 

%

\vspace{-1mm}
\section{String-Manipulating Programs with Integer Data Type}\label{sec:logic}


In this paper, we consider logics involving two data-types, i.e., the string data-type and the integer data-type. As a convention, $u, v, \dots$ denote string constants,  $c, d,\dots$ denote integer constants, $x, y, \dots$ denote string variables, and $i, j, \dots$ denote  integer variables.

We consider symbolic execution of string-manipulating programs with numeric conditions (abbreviated as {\slint}), defined by the following rules, 
\[
\begin{array}{l c l}
S &::= &  x:= y \concat z \mid x:= \replaceall_{e,u}(y) \mid   x:=\reverse(y) \mid x:=\NFT(y) \mid \\
& & x := \substring(y, t_1, t_2)  \mid \ASSERT{\varphi}  \mid S;S, \\
\varphi &::= & x \in \NFA \mid t_1\ o\ t_2 \mid \varphi \vee \varphi \mid \varphi \wedge \varphi,
\end{array}
\]
where $e$ is a regular expression over $\Sigma$, $u \in \Sigma^*$, $\NFT$ is an FFT,  $\NFA$ is an NFA, $o \in \{=, \neq, \ge, \le, >, <\}$, and $t_1,t_2$ are integer terms defined by the following rules,
\[
t  ::= i \mid c \mid \length(x) \mid \indexof_v(x, i) \mid  ct  \mid t + t, \mbox{ where } c \in \Int, v \in \Sigma^+.
\]
%
We require that the string-manipulating programs are in {\bf single static assignment (SSA) form}. Note that SSA form imposes restrictions only on the assignment statements, but not on the assertions. 
A string variable $x$ in an {\slint} program $S$ is called an \emph{input string variable} of $S$ if it does not appear on the left-hand side of the assignment statements of $S$. A variable in $S$ is called an \emph{input variable}  if it is either an input string variable or an integer variable.

\noindent\emph{Semantics.}
The semantics of {\slint} is explained as follows. 
\begin{itemize}
\item The assignment $x:=y \cdot z$ denotes that $x$ is the concatenation of two strings $y$ and $z$.

\item The assignment $x:=\replaceall_{e,u}(y)$ denotes that $x$ is the string obtained by replacing all occurrences of $e$ in $y$ with $u$, where the \emph{leftmost and longest} matching of $e$ is used. For instance, $\replaceall_{(ab)^+,c}(aababaab) =ac \cdot \replaceall_{(ab)^+,c}(aab)= acac$, since the leftmost and longest matching of $(ab)^+$ in $aababaab$ is $abab$. Here we require that the language defined by $e$ does \emph{not} contain the empty string, in order to avoid the troublesome definition of the semantics of the matching of the empty string.  The formal semantics of the $\replaceall$ function can be found in \cite{CCH+18}.
\item The assignment $x:=\reverse(y)$ denotes that $x$ is the reverse of $y$.
\item The assignment $x:=\NFT(y)$ denotes that $(y,x) \in \Tran(\NFT)$. 
\item The assignment $x:=\substring(y, t_1, t_2)$ denotes that $x$ is equal to the return value of $\substring(y, t_1, t_2)$, where 

\[ \substring(y, t_1, t_2)=
\begin{cases}
\epsilon & \mbox{if }t_1<0\vee t_1 \ge |y| \vee t_2=0 \\
y[t_1, \min\{t_1+t_2-1, |y|-1\}] & o/w
\end{cases}
\]
For instance, $\substring(abaab, -1,1)=\varepsilon$, $\substring(abaab, 3,0)=\varepsilon$, $\substring(abaab, 3,2)=ab$, and $\substring(abaab, 3,3)=ab$.
%
\item The conditional statement $\ASSERT{x \in \NFA}$ denotes that $x$ belongs to $\Lang(\NFA)$.
\item The conditional statement $\ASSERT{t_1 \ o\ t_2}$ denotes that the value of $t_1$ is equal to (not equal to, \dots) that of $t_2$, if $o\in \{ =, \neq, \geq, >, \leq, < \}$.
\item The integer term $\length(x)$ denotes the length of $x$. 
\item The function $\indexof_v(x, i)$ returns the starting position of the first occurrence of $v$ in $x$ after the position $i$, if such an occurrence exists, and $-1$ otherwise. Note that if $i<0$, then $\indexof_v(x, i)$ returns $\indexof_v(x, 0)$, and if $i \ge \length(x)$, then $\indexof_v(x, i)$ returns $-1$. For instance, $\indexof_{ab}(aaba, -1) = 1$, $\indexof_{ab}(aaba, 1) = 1$, $\indexof_{ab}(aaba, 2)=-1$, and $\indexof_{ab}(aaba, 4)=-1$.
\end{itemize}

\noindent
\textbf{Path feasibility problem.} Given an {\slint} program $S$, decide whether there are valuations of the input variables so that $S$ can execute to the end.

\vspace{-1mm}
\section{Decision Procedures for Path Feasibility}\label{sec:dec}


In this section, we present a decision procedure for the path feasibility problem of {\slint}. A distinguished feature of the decision procedure is that it conducts backward computation which is lazy and can be done in a modular way. To support this, we extend  a regular language with quantitative information of the strings in the language, giving rise to cost-enriched regular languages and corresponding finite automata (Section \ref{sect:ce}). The crux of the decision procedure is thus to show that the  pre-images of cost-enriched regular languages under the string operations in {\slint} (i.e., concatenation $\concat$, $\replaceall_{e,u}$, $\reverse$, FFTs $\NFT$, and $\substring$) are representable by so called cost-enriched recognisable relations (Section \ref{sect:pre}). The overall decision procedure is presented in Section~\ref{sec:dc}, supplied by additional complexity analysis. 

\vspace{-3mm}
\subsection{Cost-Enriched Regular Languages and Recognisable Relations} \label{sect:ce}

Let $k \in \Nat$ with $k > 0$. A \emph{$k$-cost-enriched string} is $(w, (n_1, \cdots, n_k))$ where $w$ is a string and $n_i \in \intnum$ for all $i \in [k]$. A \emph{$k$-cost-enriched language} $L$ is a subset of $\Sigma^* \times \intnum^k$. For our purpose, we identify a ``regular" fragment of cost-enriched languages as follows.  

\begin{definition}[Cost-enriched regular languages]
	Let $k \in \Nat$ with $k > 0$. A $k$-cost-enriched language is \emph{regular} (abbreviated as CERL) if it can be accepted by a \emph{cost-enriched finite automaton}. 
	
	A cost-enriched finite automaton (CEFA) $\CEFA$ is a tuple $(Q, \Sigma, R, \delta, I, F)$ where 
	\begin{itemize}
		\item $Q, \Sigma, I, F$ are defined as in NFAs, 
		\item $R=(r_1, \cdots, r_k)$ is a vector of (mutually distinct) \emph{cost registers}, 
		\item $\delta$ is the transition relation which is a finite set of tuples $(q, a, q', \eta)$ where $q, q' \in Q$, $a \in \Sigma$, and 
		$\eta: R \rightarrow \Int$
		is a cost register update function. \\
		For convenience, we usually write $(q, a, q', \eta) \in \Delta$ as $q \xrightarrow{a, \eta} q'$.
	\end{itemize}
	A \emph{run} of $\CEFA$ on a $k$-cost-enriched string $(a_1 \cdots a_m, (n_1, \cdots,n_k))$ is a  transition sequence $q_0 \xrightarrow{a_1, \eta_1} q_1 \cdots q_{m-1} \xrightarrow{a_m, \eta_m} q_m$ such that $q_0 \in I$ and $n_i = \sum \limits_{1\leq j\leq m}\eta_j(r_i)$ for each $i \in [k]$ (Note that the initial values of cost registers are zero). The run is \emph{accepting} if $q_m \in F$. A $k$-cost-enriched string $(w, (n_1, \cdots,n_k))$ is accepted by $\CEFA$ if there is an accepting run of $\CEFA$ on $(w, (n_1, \cdots,n_k))$. In particular, $(\varepsilon, n)$ is accepted by $\CEFA$ if $n=0$ and $I\cap F \neq \emptyset$.
	The $k$-cost-enriched language defined by $\CEFA$, denoted by $\Lang(\CEFA)$, is the set of $k$-cost-enriched strings accepted by $\CEFA$. 
\end{definition}
The \emph{size} of a CEFA $\CEFA=(Q, \Sigma, R, \delta, I, F)$, denoted by $|\CEFA|$, is defined as the sum of the sizes of its transitions, where the size of each transition $(q, a, q', \eta)$ is $\sum \limits_{r \in R} \lceil \log_2 (|\eta(r)|) \rceil +3$. Note here  the integer constants in $\CEFA$ are encoded in binary.

\begin{remark}
	CEFAs can be seen as a variant of Cost Register Automata \cite{RLJ+13}, by admitting nondeterminism and discarding partial final cost functions. CEFAs are also closely related to monotonic counter machines \cite{LB16}. The main difference is that CEFAs discard guards in transitions and allow binary-encoded integers in cost updates, while monotonic counter machines allow guards in transitions but restrict the cost updates to being monotonic and unary, i.e. $0,1$ only. Moreover, we explicitly define CEFAs as language acceptors for  cost-enriched languages.
\end{remark}

\begin{example}[CEFA for $\length$]\label{exm:len}
	The string function $\length$ can be captured by CEFAs. For any NFA $\NFA=(Q, \Sigma,  \delta, I, F)$, it is not difficult to see that the cost-enriched language $\{(w, \length(w)) \mid w\in \Lang(\NFA)\}$ is accepted by a CEFA, i.e., 
	$(Q, \Sigma, (r_1), \delta', I, F)$  
	such that for each $(q, a, q')\in \delta$, we let $(q, a, q', \eta)\in \delta'$, where $\eta(r_1) = 1$. 
	
	For later use, we identify a special $\CEFA_{\rm len}= (\{q_0\}, \Sigma, (r_1), \{(q_0, a, q_0, \eta) \mid \eta(r_1) = 1\}, \{q_0\}, \{q_0\})$. In other words, $\CEFA_{\rm len}$ accepts $\{(w, \length(w)) \mid w\in \Sigma^*\}$.
\end{example}

We can show that the function $\indexof_v(\cdot, \cdot)$ can be captured by a CEFA as well, in the sense that, for any NFA $\NFA$ and constant string $v$, we can construct a CEFA 
$\CEFA_{\indexof_v}$ accepting $\left\{(w, (n, \indexof_v(w, n)))\mid w\in \Lang(\NFA), n \le \indexof_v(w, n) < |w| \right\}$. 
The construction is slightly technical and can be found in \iftoggle{full}{Appendix~\ref{appendix:cefa_indexof}}{the full version \cite{atva20-full}}.

Note that $\CEFA_{\indexof_v}$ does not model the corner cases in the semantics of $\indexof_v$, for instance, $\indexof_v(w, n) = -1$ if $v$ does not occur after the position $n$ in $w$.

Given two CEFAs $\CEFA_1 = ( Q_1, \Sigma, R_1, \delta_1, I_1, F_1)$ and $\CEFA_2 = (Q_2, \Sigma, \delta_2, R_2, I_2, F_2)$ with $R_1 \cap R_2 = \emptyset$, 
the product of $\CEFA_1$ and $\CEFA_2$, denoted by $\CEFA_1 \times \CEFA_2$, is defined as $(Q_1 \times Q_2, \Sigma, R_1 \cup R_2, \delta, I_1 \times I_2, F_1 \times F_2)$, where $\delta$ comprises the tuples $((q_1, q_2), \sigma, (q'_1, q'_2), \eta)$ such that $(q_1, \sigma, q'_1, \eta_1) \in \delta_1$, $(q_2, \sigma, q'_2, \eta_2) \in \delta_2$, and $\eta = \eta_1\cup \eta_2$.  

For a CEFA $\CEFA$, we use $R(\CEFA)$ to denote the vector of cost registers occurring in $\CEFA$. 
Suppose $\CEFA$ is  CEFA with $R(\CEFA)=(r_1,\cdots, r_k)$ and $\vec{i} = (i_1,\cdots, i_k)$ is a vector of mutually distinct integer variables such that $R(\NFA) \cap \vec{i} = \emptyset$. We use $\CEFA[\vec{i}/R(\CEFA)]$ to denote the CEFA obtained from $\CEFA$ by simultaneously replacing $r_j$ with $i_j$ for $j \in [k]$. 



\begin{definition}[Cost-enriched recognisable relations]
	Let $(k_1,\cdots, k_l) \in \Nat^l$ with $k_j > 0$ for every $j \in [l]$. A cost-enriched recognisable relation (CERR)  $\cR \subseteq (\Sigma^* \times \intnum^{k_1}) \times \cdots  \times (\Sigma^* \times \intnum^{k_l})$ is a finite union of products of CERLs. Formally,
	$\cR = \bigcup \limits_{i=1}^n L_{i,1 } \times \cdots \times L_{i, l}$, 
	where for every $j \in [l]$, $L_{i,j} \subseteq \Sigma^* \times \intnum^{k_j}$ is a CERL. 
	A CEFA representation of $\cR$ is a collection of CEFA tuples $(\CEFA_{i,1}, \cdots, \CEFA_{i,l})_{i \in [n]}$ such that $\Lang(\CEFA_{i,j}) = L_{i,j}$ for every $i \in [n]$ and $j \in [l]$.
\end{definition}

\vspace{-4mm}
\subsection{Pre-images of CERLs under string operations} \label{sect:pre}

To unify the presentation, 
we consider string functions $f: (\Sigma^* \times \Int^{k_1}) \times \cdots \times (\Sigma^* \times \Int^{k_l}) \rightarrow \Sigma^*$. (If there is no integer input parameter, then $k_1,\cdots,k_l$ are zero.)  

\begin{definition}[Cost-enriched pre-images of CERLs] \label{def:preimage}
Suppose that $f: (\Sigma^* \times \Int^{k_1}) \times \cdots \times (\Sigma^* \times \Int^{k_l}) \rightarrow \Sigma^*$ is a string function, $L \subseteq \Sigma^* \times \Int^{k_0}$ is a CERL defined by a CEFA $\CEFA=(Q, \Sigma, R, \delta, I, F)$ with $R= (r_1, \cdots, r_{k_0})$. Then the $R$-cost-enriched pre-image of $L$ under $f$, denoted by $f^{-1}_R(L)$, is a pair $(\cR, \vec{t})$ such that 
\begin{itemize}
\item $\cR \subseteq (\Sigma^* \times \Int^{k_1 + k_0}) \times \cdots \times (\Sigma^* \times \Int^{k_l + k_0})$;
\item $\vec{t} = (t_1, \cdots ,t_{k_0})$ is a vector of linear integer terms where for each $i \in [k_0]$, $t_i$ is a term whose variables are from $\left\{r^{(1)}_i, \cdots, r^{(l)}_i\right\}$ which are fresh cost registers and are disjoint from $R$ in $\CEFA$;

\item $L$ is equal to the language comprising the $k_0$-cost-enriched strings
\[\left(w_0, t_1\left[d^{(1)}_{1}/r^{(1)}_1, \cdots, d^{(l)}_{1}/r^{(l)}_1\right], \cdots, t_{k_0}\left[d^{(1)}_{k_0}/r^{(1)}_{k_0}, \cdots, d^{(l)}_{k_0}/r^{(l)}_{k_0}\right]
\right), \]
such that 
\[w_0 = f\left((w_1, \vec{c_1}), \cdots, (w_l, \vec{c_l}\right)) \mbox{ for some } ((w_1, (\vec{c_1}, \vec{d_1})), \cdots, (w_l, (\vec{c_l}, \vec{d_l}))) \in \cR,\]
where $\vec{c_j} \in \Int^{k_j}$, $\vec{d_j} = (d^{(j)}_{1}, \cdots, d^{(j)}_{k_0}) \in \Int^{k_0}$ for $j\in [l]$.
%
\end{itemize}
The $R$-cost-enriched pre-image of $L$ under $f$, say $f^{-1}_R(L)=(\cR, \vec{t})$, is said to be CERR-definable if $\cR$ is a CERR. 
\end{definition}

Definition~\ref{def:preimage} is essentially a semantic definition of the pre-images. For the decision procedure, one desires an effective representation of a CERR-definable $f^{-1}_R(L)=(\cR, \vec{t})$ in terms of CEFAs. Namely,
a CEFA representation of 
$(\cR, \vec{t})$ (where $t_j$ is over $\left\{r^{(1)}_j, \cdots, r^{(l)}_j\right\}$ for $j\in [k_0]$)
is a tuple $((\CEFA_{i,1}, \cdots, \CEFA_{i, l})_{i \in [n]}, \vec{t})$ such that $(\CEFA_{i,1}, \cdots, \CEFA_{i, l})_{i \in [n]}$ is a CEFA representation of $\cR$, where $R(\CEFA_{i,j})=\left(r'_{j,1}, \cdots, r'_{j,k_j}, r^{(j)}_1, \cdots,r^{(j)}_{k_0}\right)$ for each $i \in [n]$ and $j \in [l]$. (The cost registers $r'_{1,1}, \cdots, r'_{1,k_1},\cdots, r'_{l,1}, \cdots, r'_{l,k_l}$ 
are mutually distinct and freshly introduced.) 

\begin{example}[$\substring^{-1}_R(L)$]\label{exm:pre-image}
Let $\Sigma = \{a\}$ and $L = \{(w, |w|) \mid w \in \Lang((aa)^*) \}$. Evidently $L$  is a CERL defined by a CEFA $\CEFA = (Q, \Sigma, R, \delta, \{q_0\}, \{q_0\})$ with $Q=\{q_0,q_1\}$, $R=(r_1)$ and $\delta = \{(q_0, a, q_1), (q_1, a, q_0)\}$. Since $\substring$  is  from $\Sigma^* \times \Int^2$ to $\Sigma^*$, $\substring^{-1}_R(L)$, the $R$-cost-enriched pre-image of $L$ under $\substring$, is the pair $(\cR, t)$, where $t=r^{(1)}_1$ (note that in this case $l=1$, $k_0=1$, and $k_1=2$) and 
$$\cR = \{(w, n_1, n_2, n_2) \mid w \in \Lang(a^*), n_1 \ge 0, n_2 \ge 0, n_1+n_2 \le |w|, n_2 \mbox{ is even}\},$$ 
which is represented by $(\CEFA', t)$ such that $\CEFA'= (Q', \Sigma, R', \delta', I', F')$, where 
\begin{itemize}
\item $Q' = Q \times \{p_0, p_1, p_2\}$, (Intuitively, $p_0$, $p_1$, and $p_2$ denote that the current position is before the starting position, between the starting position and ending position, and after the ending position of the substring respectively.) 
\item $R'= \left(r'_{1,1}, r'_{1,2}, r^{(1)}_1 \right)$, 
\item $I' =\{(q_0,p_0)\}$, $F'=\{(q_0, p_2), (q_0, p_0)\}$ (where $(q_0, p_0)$ is used to accept the $3$-cost-enriched strings $(w, n_1, 0, 0)$ with $0 \le n_1 \le |w|$), and 
\item $\delta'$ is  
\[
\left\{
\begin{array}{l}
(q_0, p_0) \xrightarrow{a, \eta_1} (q_0, p_0), (q_0, p_0) \xrightarrow{a, \eta_2} (q_1, p_1), (q_1, p_1) \xrightarrow{a, \eta_2} (q_0, p_1), \\
(q_0, p_1) \xrightarrow{a, \eta_2} (q_1, p_1), (q_1, p_1) \xrightarrow{a, \eta_2} (q_0, p_2), (q_0, p_2) \xrightarrow{a, \eta_3} (q_0, p_2)
\end{array}
\right\},
\] 
where $\eta_1(r'_{1,1})=1$, $\eta_1(r'_{1,2})=0$, $\eta_1(r^{(1)}_1)=0$, $\eta_2(r'_{1,1})=0$, $\eta_2(r'_{1,2})=1$, and $\eta_2(r^{(1)}_1)=1$, $\eta_3(r'_{1,1})=0$, $\eta_3(r'_{1,2})=0$, and $\eta_3(r^{(1)}_1)=0$.
\end{itemize}
Therefore, $\substring^{-1}_R(L)$ is CERR-definable.
\end{example}

It turns out that for each string function $f$ in the assignment statements of {\slint}, the cost-enriched pre-images of CERLs under $f$ are CERR-definable.

\begin{proposition}\label{prop:pre-image}
Let $L$ be a CERL defined by a CEFA $\CEFA = (Q, \Sigma, R, \delta, I, F)$. Then for each string function $f$ ranging over $\concat$, $\replaceall_{e,u}$, $\reverse$, FFTs $\NFT$, and $\substring$, $f^{-1}_R(L)$ is CERR-definable. In addition,
\begin{itemize}
\item a CEFA representation of $\concat^{-1}_R(L)$ can be computed in time $\bigO(|\CEFA|^2)$, 
\item a CEFA representation of $\reverse^{-1}_R(L)$ (resp. $\substring^{-1}_R(L)$) can be computed in time $\bigO(|\CEFA|)$,
\item a CEFA representation of  $(\Tran(\NFT))^{-1}_R(L)$ can be computed in time polynomial in $|\CEFA|$ and exponential in $|\NFT|$,
\item a CEFA representation of  $(\replaceall_{e,u})^{-1}_R(L)$ can be computed in time polynomial in $|\CEFA|$ and exponential in $|e|$ and $|u|$.
\end{itemize}
\end{proposition}

The proof of Proposition~\ref{prop:pre-image} is given in \iftoggle{full}{Appendix~\ref{app:pre-image}}{the full version \cite{atva20-full}}.

\vspace{-2mm}
\subsection{The Decision Procedure}\label{sec:dc}
%
Let $S$  be an {\slint} program. 
Without loss of generality, we assume that for every occurrence of assignments of the form $y:= \substring(x, t_1, t_2)$, it holds that $t_1$ and $t_2$ are integer variables. This is not really a restriction, since, for instance, if in $y:= \substring(x, t_1, t_2)$, neither $t_1$ nor $t_2$ is an integer variable, then we introduce fresh integer variables $i$ and $j$, replace $t_1, t_2$ by $i,j$ respectively, and add $\ASSERT{i=t_1};\ASSERT{j = t_2}$ in $S$.
We present a decision procedure for the path feasibility problem of $S$ which is divided into five steps. 
 
\smallskip
\noindent {\bf Step I: Reducing to atomic assertions.}

Note first that in our language, each assertion is a positive Boolean combination of atomic formulas of the form $x\in \CEFA$ or $t_1\ o\ t_2$ (cf. Section~\ref{sec:logic}). Nondeterministically choose, for each assertion $\ASSERT{\varphi}$ of $S$, a set of atomic formulas $\Phi_\varphi = \{\alpha_1,\cdots,\alpha_n\}$ such that $\varphi$ holds when atomic formulas in $\Phi_\varphi$ are true.  

Then each assertion $\ASSERT{\varphi}$ in $S$ with $\Phi_\varphi = \{\alpha_1,\cdots,\alpha_n\}$ is replaced by $\ASSERT{\alpha_1}; \cdots; \ASSERT{\alpha_n}$, and thus $S$ constrains atomic assertions only. 

\smallskip 
\noindent {\bf Step II: Dealing with the case splits in the semantics of $\indexof_v$ and $\substring$.}


For each integer term of the form $\indexof_v(x,i)$ in $S$, nondeterministically choose one of the following five options (which correspond to the semantics of $\indexof_v$ in Section~\ref{sec:logic}).
\begin{itemize}
\item[(1)] Add $\ASSERT{i < 0}$ to $S$, and replace $\indexof_v(x,i)$ with $\indexof_v(x,0)$ in $S$. 
\item[(2)] Add $\ASSERT{i < 0};\ASSERT{x \in \NFA_{\overline{\Sigma^*v\Sigma^*}}}$ to $S$; replace $\indexof_v(x,i)$ with $-1$ in $S$.
\item[(3)] Add $\ASSERT{i \ge \length(x)}$ to $S$, and replace $\indexof_v(x,i)$ with $-1$ in $S$.
\item[(4)] Add $\ASSERT{i \ge 0}; \ASSERT{i < \length(x)}$ to $S$.
\item[(5)] Add 
$$
\begin{array}{l}
\ASSERT{i \ge 0}; \ASSERT{i < \length(x)}; \ASSERT{j=\length(x)-i}; \\
\ \ \ \ y:=\substring(x, i, j); \ASSERT{y \in \NFA_{\overline{\Sigma^*v\Sigma^*}}}
\end{array}
$$ 
to $S$, where $y$ is a fresh string variable, $j$ is a fresh integer variable, and $\NFA_{\overline{\Sigma^*v\Sigma^*}}$ is an NFA defining the language $\{w \in \Sigma^*\mid v \mbox{ does not occur as a substring in } w\}$. Replace $\indexof_v(x, i)$ with $-1$ in $S$.
\end{itemize}

For each assignment $y:=\substring(x, i, j)$, nondeterministically choose one of the following three options (which correspond to the semantics of $\substring$ in Section~\ref{sec:logic}).
\begin{itemize}
\item[(1)] Add the statements $\ASSERT{i \ge 0}; \ASSERT{i + j \le \length(x)}$ to $S$. 
\item[(2)] Add the statements 
$\ASSERT{i \ge 0}; \ASSERT{i \le \length(x)};\ASSERT{i+j  > \length(x)}$; $\ASSERT{i'  = \length(x)-i}$
to $S$, and replace $y:=\substring(x, i, j)$ with $y:=\substring(x, i, i')$, where $i'$ is a fresh integer variable.
\item[(3)] Add the statement $\ASSERT{i < 0}; \ASSERT{y \in \NFA_\varepsilon}$ to $S$, and  remove $y:=\substring(x, i, j)$ from $S$, where $\NFA_\varepsilon$ is the NFA defining the language $\{\varepsilon\}$.
\end{itemize}

\smallskip
\noindent {\bf Step III: Removing $\length$ and $\indexof$}.



For each term $\length(x)$ in $S$, we introduce a \emph{fresh} integer variable $i$, replace every occurrence of $\length(x)$ by $i$, and add the statement $\ASSERT{x \in \CEFA_{\rm len}[i/r_1]}$ to $S$. (See Example~\ref{exm:len} for the definition of $\CEFA_{\rm len}$.)  

For each term $\indexof_v(x, i)$ occurring in $S$, introduce two fresh integer variables $i_1$ and $i_2$, replace every occurrence of $\indexof_v(x, i)$ by $i_2$, and add the statements $\ASSERT{i=i_1}; \ASSERT{x \in \CEFA_{\indexof_v}[i_1/r_1, i_2/r_2]}$ to $S$.  

 \smallskip
\noindent {\bf Step IV: Removing the assignment statements backwards}.


Repeat the following procedure until $S$ contains no assignment statements.
\begin{quote}
Suppose $y := f(x_1, \vec{i_1}, \cdots, x_l, \vec{i_l})$ is the \emph{last} assignment of $S$, where $f: (\Sigma^* \times \Int^{k_1}) \times \cdots \times (\Sigma^* \times \Int^{k_l}) \rightarrow \Sigma^*$ is a string function and $\vec{i_j}= (i_{j,1}, \cdots, i_{j, k_j})$ for each $j \in [l]$.
\\
Let $\{\CEFA_1, \cdots, \CEFA_s\}$ be the set of all CEFAs such that $\ASSERT{y \in \CEFA_j}$ occurs in $S$ for every $j \in [s]$. 
Let $j \in [s]$ and $R(\CEFA_j)=(r_{j,1}, \cdots, r_{j, \ell_j})$. Then from Proposition~\ref{prop:pre-image}, 
a CEFA representation of $f^{-1}_{R(\CEFA_j)}(\Lang(\CEFA_j))$, say $\left(\left(\cB^{(1)}_{j, j'}, \cdots, \cB^{(l)}_{j, j'}\right)_{j' \in [m_j]}, \vec{t}\right)$, can be effectively computed from $\NFA$ and $f$, where we write
\[
R\left(\cB^{(j'')}_{j, j'}\right)=\left((r')^{(j'',1)}_{j}, \cdots, (r')_{j}^{(j'',k_{j''})}, r^{(j'')}_{j, 1}, \cdots,r^{(j'')}_{j, \ell_j} \right)
\]
for each $j' \in [m_j]$ and $j'' \in [l]$, and $\vec{t}=(t_1,\cdots, t_{\ell_j})$. Note that the cost registers $(r')^{(1,1)}_{j}, \cdots, (r')_{j}^{(1,k_1)}, \cdots, (r')^{(l,1)}_{j}, \cdots, (r')_{j}^{(l,k_l)}, r^{(1)}_{j, 1}, \cdots,r^{(1)}_{j, \ell_j}, \cdots, r^{(l)}_{j, 1}, \cdots,r^{(l)}_{j, \ell_j}$ are mutually distinct and freshly introduced, moreover, $R\left(\cB^{(j'')}_{j, j'_1}\right)=R\left(\cB^{(j'')}_{j, j'_2}\right)$ for distinct $j'_1,j'_2 \in [m_j]$.

Remove $y := f(x_1, \vec{i_1}, \cdots, x_l, \vec{i_l})$, as well as all the statements $\ASSERT{y \in \CEFA_1}$, $\cdots$, $\ASSERT{y \in \CEFA_s}$ from $S$. For every $j \in [s]$, nondeterministically choose $j' \in [m_j]$, and add the following statements to $S$, 
\[
\begin{array}{l}
\ASSERT{x_1 \in \cB^{(1)}_{j, j'}};\ \cdots;\ \ASSERT{x_l \in \cB^{(l)}_{j, j'}}; S_{j, j', \vec{i_1}, \cdots, \vec{i_l}}; S_{j, \vec{t}}\\
\end{array}
\]
where 
\[
\begin{array}{l c c}
S_{j, j', \vec{i_1}, \cdots, \vec{i_l}} & \equiv & \ASSERT{i_{1,1} = (r')^{(1,1)}_{j, j'}}; \cdots; \ASSERT{i_{1,k_1} = (r')^{(1,k_1)}_{j, j'}};\\
& & \cdots\\
 & & \ASSERT{i_{l,1} = (r')^{(l,1)}_{j, j'}}; \cdots; \ASSERT{i_{l,k_l} = (r')^{(l,k_l)}_{j, j'}}
\end{array}
\]
and
\[
\begin{array}{l}
S_{j, \vec{t}} \equiv \ASSERT{r_{j, 1} = t_1}; \cdots, \ASSERT{r_{j, \ell_j} = t_{\ell_j}}.
\end{array}
\]
\end{quote}

\smallskip
\noindent{\bf Step V: Final satisfiability checking.} 


In this step, $S$ 
contains no assignment statements and only assertions of the form $\ASSERT{x \in \CEFA}$ and  $\ASSERT{t_1\ o\ t_2}$  where $\CEFA$ are CEFAs and $t_1, t_2$ are linear integer terms. 
Let $X$ denote the set of string variables occurring in $S$.
For each $x \in X$, let $\Lambda_x=\{\CEFA_{x}^1, \cdots, \CEFA_{x}^{s_x}\}$ denote the set of CEFAs $\CEFA$ such that $\ASSERT{x \in \CEFA}$ appears in $S$. 
Moreover, let $\phi$ denote the conjunction of all the LIA formulas $t_1\ o\ t_2$ occurring in $S$. It is straightforward to observe that $\phi$ is over 
$R'=\bigcup_{x\in X, j \in [s_x]}R(\CEFA_{x}^{j})$. Then the path feasibility of $S$ is reduced to \emph{the satisfiability problem of LIA formulas w.r.t. CEFAs (abbreviated as {\lasat} problem)} which is defined as 
\begin{quote}
deciding whether $\phi$ is satisfiable w.r.t. $(\Lambda_x)_{x \in X}$, namely, 
whether there are an assignment function $\theta: R' \rightarrow \Int$ and strings $(w_x)_{x \in X}$ such that  $\phi[\theta(R')/R']$ holds and $(w_x, \theta(R(\CEFA_{x}^{j}))) \in \Lang(\CEFA_{x}^{j})$ for every $x \in X$ and $j \in [s_x]$.
\end{quote}
This {\lasat} problem is decidable and $\pspace$-complete;
The proof can be found in \iftoggle{full}{Appendix~\ref{app:sat-cefa}}{the full version \cite{atva20-full}}.

%
%

\begin{proposition}\label{prop:la-sat-cefa-inter}
{\lasat} is $\pspace$-complete.
\end{proposition}
%
%
An example to illustrate the decision procedure can be found in \iftoggle{full}{Appendix~\ref{app:urlexample}}{the full version \cite{atva20-full}}.

\smallskip
\noindent\emph{Complexity analysis of the decision procedure.} Step I and Step II can be done in nondeterministic linear time. Step III can be done in linear time. In Step IV, for each input string variable $x$ in $S$, at most exponentially many CEFAs can be generated for $x$, each of which is of at most exponential size. Therefore, Step IV can be done in nondeterministic exponential space. By Proposition~\ref{prop:la-sat-cefa-inter}, Step V can be done in exponential space. Therefore, we conclude that the path feasibility problem of {\slint} programs is in $\nexpspace$, thus in $\expspace$ by Savitch's theorem \cite{complexity-book}.  
 
\begin{remark}
	In this paper, we focus on functional finite transducers (cf.\ Section~\ref{sec:prel}). Our decision procedure is applicable to general finite transducers as well with minor adaptation. However, the $\expspace$ complexity upper-bound does not hold any more, because the distributive property $f^{-1}(L_1\cap L_2)= f^{-1}(L_1)\cap f^{-1}(L_2)$ for regular languages $L_1, L_2$ only holds for functional finite transducers $f$.  
\end{remark}

\vspace{-3mm}
\section{Evaluations} \label{sec:eva}


We have implemented the decision procedure presented in the preceding section based on the recent string constraint solver \ostrich~\cite{CHL+19}, resulting in a new solver \ostrich+. \ostrich\ is  written in Scala and based on the SMT solver Princess \cite{princess08}. \ostrich+ reuses the parser of Princess, but replaces the NFAs from \ostrich\ with CEFAs. Correspondingly, in \ostrich+, the pre-image  computation for concatenation, $\replaceall$, $\reverse$, and finite transducers is reimplemented, and a new pre-image operator for $\substring$ is added. \ostrich+ also implements CEFA constructions for $\length$ and $\indexof$.  \iftoggle{full}{More details can be found in Appendix~\ref{appendix:impl}}{More details can be found in the full version~\cite{atva20-full}}.

We have compared {\ostrich}+ with some of the state-of-the-art solvers on a wide range of benchmarks.  
We  discuss the benchmarks in Section~\ref{sec:bench} and 
present the experimental results in Section~\ref{sec:exp-res}.

\vspace{-3mm}
\subsection{Benchmarks}\label{sec:bench}
Our evaluation focuses on problems that combine string with integer constraints.  To this end, we consider the following four sets of
benchmarks, all in SMT-LIB~2 format.

\smallskip
\noindent \transducerbench+
is derived from the {\transducerbench} benchmark suite of {\ostrich}
\cite{CHL+19}.  The {\transducerbench} suite involves seven
transducers: toUpper (replacing all lowercase letters with their
uppercase ones) and its dual toLower, htmlEscape   and
its dual htmlUnescape, escapeString, addslashes, and trim. 
These transducers are
collected from Stranger \cite{YABI14} and SLOTH
\cite{HJLRV18}. Initially none of the benchmarks involved integers. In
{\transducerbench+}, we encode four security-relevant properties of
transducers~\cite{BEK}, with the help of the functions~$\charat$ and
$\length$:
\begin{itemize}
\item idempotence: given $\NFT$, whether
  $\forall x.\ \NFT(\NFT(x)) = \NFT(x)$;
\item duality: given $\NFT_1$ and
  $\NFT_2$, whether $\forall x.\ \NFT_2(\NFT_1(x)) = x$;
\item commutativity: given $\NFT_1$ and $\NFT_2$, whether
  $\forall x.\ \NFT_2(\NFT_1(x)) = \NFT_1(\NFT_2(x))$;
\item equivalence: given $\NFT_1$ and $\NFT_2$, whether
  $\forall x.\ \NFT_1(x) = \NFT_2(x)$.
\end{itemize}

%
For instance, we encode the non-idempotence of $\NFT$ into the path feasibility of the {\slint} program $y:=\NFT(x); z:=\NFT(y); S_{y \neq z}$, where $y$ and $z$ are two fresh string variables, and $S_{y \neq z}$ is the {\slint} program encoding $y \neq z$ (see 
\iftoggle{full}{Appendix~\ref{appendix:slint-prog-ineq} for the details}{the full version \cite{atva20-full} for the details}
). We also include in {\transducerbench+} three instances 
generated from a program to sanitize URLs against XSS attacks (see \iftoggle{full}{Appendix~\ref{exmp:running} for the details}{the full version \cite{atva20-full} for the details}), 
where $\NFT_{\rm trim}$ is used. 
In total, we obtain 94 instances for the {\transducerbench+} suite. 

\smallskip
\noindent{\slogbench} is adapted from the SLOG benchmark suite
~\cite{fang-yu-circuits}, containing 3,511~instances about strings only.
We  obtain  {\slogbench}  by choosing a string variable $x$ for each instance,
and adding the statement $\ASSERT{\length(x) < 2\ \indexof_{a}(x, 0)}$ for some $a \in \Sigma$.
As in \cite{CHL+19}, we  split  {\slogbench}  into  \slogbenchr\ and \slogbenchra,  comprising 3,391 and 120 instances respectively. In addition to 
the $\indexof$ and $\length$ functions, the benchmarks use regular constraints and concatenation;  {\slogbenchr} also contains the $\replace$ function (replacing the first occurrence), while {\slogbenchra}  uses the $\replaceall$ function (replacing all occurrences).

\smallskip
\noindent \pyexbench~\cite{ReynoldsWBBLT17} 
contains 25,421 instances  derived by the PyEx tool, a symbolic execution engine for Python programs. The {\pyexbench} suite was generated by the CVC4 group from four popular Python packages: httplib2, pip, pymongo, and requests. These instances use regular constraints, concatenation, $\length$, $\substring$, and $\indexof$ functions. Following \cite{ReynoldsWBBLT17}, the {\pyexbench} suite is further divided into three parts: {\pyextdbench},  {\pyexztbench} and {\pyexzzbench}, comprising 5,569, 8,414 and 11,438  instances, respectively. 

\smallskip
\noindent\kaluzabench~\cite{Berkeley-JavaScript}
is the most well-known  benchmark suite in  literature, 
containing 47,284 instances with regular constraints, concatenation, and the $\length$ function. The 47,284 benchmarks include 28,032 satisfiable and 9,058 unsatisfiable problems in SSA form.

\newcommand{\ostrichi}{\ostrich$^{(1)}$}
\newcommand{\ostrichii}{\ostrich$^{(2)}$}

\definecolor{Gray}{gray}{0.9}
\begin{table}[tbp]
\newcommand{\summary}[2]{\multirow{3}{*}{\begin{tabular}{c}#1\\Total: #2\end{tabular}}}
\begin{center}
\begin{tabular}{|c|c|*{6}{c|}}
\hline
Benchmark & Output &  \cvc & \zthree &  \zthreetrau & \ostrichi & \ostrichii & \ostrich+ \\
\hline
\hline
\summary{\transducerbench+}{94} & \cellcolor{Gray} sat &  \cellcolor{Gray}$-$ & \cellcolor{Gray}$-$ & \cellcolor{Gray}$-$ & \cellcolor{Gray}0 & \cellcolor{Gray}0 & \cellcolor{Gray}\bf{84}\\
\cline{2-8}
 & unsat &$-$  &$-$ &$-$ & 1 & 1 &\bf{4}\\
\cline{2-8}
 & \cellcolor{Gray}  inconcl.  &\cellcolor{Gray}$-$    &\cellcolor{Gray}$-$  &\cellcolor{Gray}$-$  &  \cellcolor{Gray}93 &  \cellcolor{Gray}93 &\cellcolor{Gray}6\\
\hline
\hline
\summary{\slogbenchra}{120} & \cellcolor{Gray} sat &  \cellcolor{Gray}\bf{104}  & \cellcolor{Gray}$-$ & \cellcolor{Gray}$-$  & \cellcolor{Gray}0 & \cellcolor{Gray}0 &98 \cellcolor{Gray}\\
\cline{2-8}
 & unsat &11  &$-$  &$-$ & 7 & 5 &\bf{12}\\
\cline{2-8}
 &\cellcolor{Gray} inconcl. & \cellcolor{Gray}5  &\cellcolor{Gray}$-$ &\cellcolor{Gray}$-$ & \cellcolor{Gray}113 & \cellcolor{Gray}115 &\cellcolor{Gray}10\\
\hline
\hline
\summary{\slogbenchr}{3,391} & \cellcolor{Gray} sat &  \cellcolor{Gray}\bf{1,309} & \cellcolor{Gray}878 & \cellcolor{Gray}$-$ & \cellcolor{Gray}0 & \cellcolor{Gray}169 & \cellcolor{Gray}584 \\
\cline{2-8}
 & unsat & \bf{2,082} & 2,066  &$-$ &2,079 & 2,075 &\bf{2,082}\\
\cline{2-8}
 &\cellcolor{Gray}  inconcl. & \cellcolor{Gray}0  &  \cellcolor{Gray}447   &  \cellcolor{Gray}$-$ & \cellcolor{Gray}1,312 & \cellcolor{Gray}1,147 &\cellcolor{Gray}725\\
\hline
\hline
\summary{\pyextdbench}{5,569} & \cellcolor{Gray} sat & \cellcolor{Gray}4,224 & \cellcolor{Gray}4,068 &  \cellcolor{Gray} \bf{4,266} & \cellcolor{Gray}68 & \cellcolor{Gray}96 & \cellcolor{Gray}4,141\\
\cline{2-8}
 & unsat & 1,284 & 1,289 & \bf{1,295} & 95 & 93 &1,203\\
\cline{2-8}
 &\cellcolor{Gray} inconcl. &\cellcolor{Gray}61 &\cellcolor{Gray}212   &\cellcolor{Gray}8 & \cellcolor{Gray}5,406 & \cellcolor{Gray}5,380 &\cellcolor{Gray}225\\
\hline
\hline
\summary{\pyexztbench}{8,414} & \cellcolor{Gray} sat & \cellcolor{Gray}6,346 & \cellcolor{Gray}6,040 & \cellcolor{Gray}\bf{7,003} & \cellcolor{Gray}76 & \cellcolor{Gray}100 & \cellcolor{Gray}5,489\\
\cline{2-8}
 & unsat & 1,358  & 1,370  &\bf{1,394} & 61 & 53 &1,239\\
\cline{2-8}
 & \cellcolor{Gray}inconcl. &\cellcolor{Gray}710 &\cellcolor{Gray}1,004 &\cellcolor{Gray} 17 & \cellcolor{Gray}8,277 & \cellcolor{Gray}8,261 &\cellcolor{Gray}1,686\\
\hline
\hline
\summary{\pyexzzbench}{11,438} & \cellcolor{Gray} sat & \cellcolor{Gray} 10,078 & \cellcolor{Gray} 8,804 & \cellcolor{Gray} \bf{10,129} & \cellcolor{Gray}71 & \cellcolor{Gray}98 & \cellcolor{Gray}9,033\\
\cline{2-8}
 & unsat & 1,204 & 1,207  &   \bf{1,222} & 91 & 61 &868\\
\cline{2-8}
 &\cellcolor{Gray}  inconcl. &\cellcolor{Gray}156 & \cellcolor{Gray}1,427  &  \cellcolor{Gray} 87 & \cellcolor{Gray}11,276 & \cellcolor{Gray}11,279 &\cellcolor{Gray}1,537 \\
\hline
\hline
\summary{\kaluzabench}{47,284} & \cellcolor{Gray} sat &  \cellcolor{Gray} \bf{35,264} & \cellcolor{Gray} 33,438 & \cellcolor{Gray} 34,769 & \cellcolor{Gray}23,397 & \cellcolor{Gray}28,522 & \cellcolor{Gray}27,962\\
\cline{2-8}
 & unsat & \bf{12,014} &  11,799  &\bf{12,014}  & 10,445 & 10,445 &9,058\\
\cline{2-8}
 &\cellcolor{Gray} inconcl. &\cellcolor{Gray}6 & \cellcolor{Gray}2,047  &\cellcolor{Gray}501 & \cellcolor{Gray}13,442 & \cellcolor{Gray}8,317 &\cellcolor{Gray}10,264 \\
\hline 
\hline
\multirow{2}{*}{Total: 76,310} & \cellcolor{Gray} solved & \cellcolor{Gray}\bf{75,278}  & \cellcolor{Gray}70,959 & \cellcolor{Gray}72,092 & \cellcolor{Gray}36,391 & \cellcolor{Gray}41,718 & \cellcolor{Gray}61,857\\
\cline{2-8}
 &  unsolved &1,032  & 5,351  & 4,218 & 39,919 & 34,592 &14,453  \\
\hline
\end{tabular}
\end{center}
\caption{Experimental results on different benchmark suites.  '--' means that the tool is not applicable to the benchmark suite, and 'inconclusive' means that a tool gave up, timed out, or crashed.}
\label{tab-experiment}\vspace{-6mm}
\end{table}%

\vspace{-3mm}
\subsection{Experiments}\label{sec:exp-res}

We compare {\ostrich}+ to {\cvc}~\cite{cvc4}, {\zthree}~\cite{Z3-str},
and {\zthreetrau} \cite{Z3-trau}, as well as two configurations of
\ostrich~\cite{CHL+19} with standard NFAs. The configuration \ostrichi\ is
a direct implementation of the algorithm in \cite{CHL+19}, and does
not support integer functions. In \ostrichii, we integrated support
for the $\length$ function as in Norn~\cite{Abdulla14}, based on the
computation of length abstractions of regular languages, and handle
$\indexof$, $\substring$, and $\charat$ via an encoding to word equations.
The experiments are executed on a computer with an Intel Xeon Silver 4210 2.20GHz and 2.19GHz CPU (2-core) and 8GB main memory, running 64bit Ubuntu 18.04 LTS OS and Java 1.8. We use a timeout of 30~seconds (wall-clock time), and report the number of satisfiable and unsatisfiable problems solved by each of the systems. Table~\ref{tab-experiment} summarises the experimental results. We did not observe incorrect answers by any tool.

There are two additional state-of-the-art solvers  {\slent} and {\trauplus} which were not included in
the evaluation. We exclude {\slent}~\cite{WC+18} because it uses its own input format laut, which is different from the SMT-LIB~2 format used for our
benchmarks; also, {\transducerbench+} is beyond the scope of {\slent}.
{\trauplus}~\cite{AbdullaA+19}  integrates {\trau} with {\sloth} to deal with both finite transducers and integer constraints. We were unfortunately unable
to obtain a working version of {\trauplus}, possibly because {\trau} requires two separate versions of Z3 to run. In addition, the algorithm in~\cite{AbdullaA+19} focuses on length-preserving transducers, which means that {\transducerbench}+ is beyond the scope of \trauplus.
 
{\ostrich}+ and \ostrich\ are the only tools applicable to the
problems in {\transducerbench}+. With a timeout of 30s, \ostrich+ can
solve 88 of the benchmarks, but this number rises to 94 when using a
longer timeout of 600s. Given the complexity of those benchmarks, this
is an encouraging result. \ostrich\ can only solve one of the
benchmarks, because the encoding of $\charat$ in the benchmarks using
equations almost always leads to problems that are not in SSA form.

On {\slogbenchra}, {\ostrich}+ and {\cvc} are very close: {\ostrich}+ solves 98 satisfiable instances, slightly less than the 104 instances solved by {\cvc}, while {\ostrich}+ solves one more unsatisfiable instance than {\cvc} (12 versus 11). The suite is beyond the scope of {\zthree} and {\zthreetrau}, which do not support $\replaceall$.

On {\slogbenchr}, {\ostrich}+, {\cvc}, and {\zthree} solve a similar
number of unsatisfiable problems, while {\cvc} solves the largest
number of satisfiable instances (1,309). The  suite 
is beyond the scope of {\zthreetrau} which does not support
$\replace$.

On the three \pyexbench\ suites, {\zthreetrau} consistently solves the
largest number of instances by some margin. \ostrich+ solves a similar
number of instances as \zthree. Interpreting the results, however, it
has to be taken into account that \pyexbench\  includes 1,334  instances
that are \emph{not} in SSA form, which are beyond the scope of
\ostrich+.

%
%

The {\kaluzabench} problems can be solved most effectively by {\cvc}. \ostrich+ can solve almost all of the around 80\% of the benchmarks
that are in SSA form, however.
%

\ostrich+ consistently outperforms \ostrichi\ and \ostrichii\ in the
evaluation, except for the \kaluzabench\ benchmarks. For
\ostrichi, this is expected because most benchmarks considered
here contain integer functions. For \ostrichii, it turns out that the
encoding of $\indexof$, $\substring$, and $\charat$ as word equations
usually leads to problems that are not in SSA form, and therefore are
beyond the scope of \ostrich.


In summary, we  observe that \ostrich+ is competitive with other solvers, while is able to handle benchmarks that are beyond
the scope of the other tools due to the combination of string functions (in particular transducers) and integer
constraints. Interestingly, the experiments show that \ostrich+, at
least in its current state, is better at solving unsatisfiable problems than satisfiable problems; this might be an artefact of the
use of nuXmv for analysing products of CEFAs. We expect that further
optimisation of our algorithm will lead to additional performance improvements. 
For instance, a natural optimisation that is to be
included in our implementation is to use standard finite automata, 
as opposed to CEFAs, for simpler problems such as the
\kaluzabench\ benchmarks. Such a combination of automata
representations is mostly an engineering effort.

%

\vspace{-3mm}
\section{Conclusion} \label{sec:conc}

In this paper, we have proposed  an expressive string constraint language which can specify constraints on both strings and integers.  We provided an automata-theoretic decision procedure for the path feasibility problem of this language. The decision procedure is simple, generic, and amenable to implementation, giving rise to a new solver OSTRICH+.  We have evaluated OSTRICH+ on  a wide range of existing and newly created benchmarks, and have obtained very encouraging results.  OSTRICH+ is shown to be the first solver  which is capable of tackling finite transducers and integer constraints with completeness guarantees. Meanwhile, it demonstrates competitive performance against some of the best state-of-the-art string constraint solvers.

\medskip
\small{
\noindent \emph{Acknowledgements.}  
T.~Chen and Z.~Wu are supported by Guangdong Science and Technology Department grant (No.\ 2018B010107004); T. Chen is also supported by Overseas Grant (KFKT2018A16) from the State Key Laboratory of
Novel Software Technology, Nanjing University, China and Natural Science Foundation of Guangdong Province, China (No. 2019A1515011689).
M.~Hague is supported by EPSRC [EP/T00021X/1];.
A.~Lin is supported by the European Research Council (ERC) under the European
Union's Horizon 2020 research and innovation programme (grant agreement no
759969).
P.~R\"ummer is supported by the
Swedish Research Council (VR) under grant 2018-04727, and by the
Swedish Foundation for Strategic Research (SSF) under the project
WebSec (Ref.\ RIT17-0011).
Z.~Wu is partially supported by  the Open Project of Shanghai Key Laboratory of Trustworthy Computing (No. 07dz22304201601), the NSFC grants (No. 61872340), and the INRIA-CAS joint research project VIP.   
}

\vspace{-3mm}
\bibliographystyle{abbrv}
\bibliography{string-short}


\iftoggle{full}
{
\newpage
\begin{appendix}

%

\section{The {\slint} program $S_{x \neq y}$ encoding $x \neq y$} \label{appendix:slint-prog-ineq}

At first, we note that the function $\charat(x, i)$ which returns $x[i]$ (i.e., the character of $x$ at the position $i$) can be seen as a special case of $\substring$, namely $\charat(x, i) \equiv \substring(x, i, 1)$. Then the string inequality $x \neq y$ is expressed as the following {\slint} program (denoted by $S_{x \neq y}$)
\[
\begin{array}{l}
z_1:=\charat(x,i); z_2 := \charat(y,i); \\
\ASSERT{\length(x) \neq \length(y) \vee \bigvee_{a \in \Sigma} (z_1 \in \NFA_a \wedge z_2 \in \NFA_{\Sigma \setminus a})},
\end{array}
\] 
where $z_1,z_2$ are two freshly introduced string variables, and $\NFA_a$ (resp. $\NFA_{\Sigma \setminus a}$) is the NFA accepting $\{a\}$ (resp. $\Sigma \setminus \{a\}$). Intuitively, two strings are different if their lengths are different or otherwise, there exists some position where the characters of the two strings are different.


\section{Construction of $\CEFA_{\indexof_v}$} \label{appendix:cefa_indexof}

In this section, we show that the function $\indexof_v(\cdot, \cdot)$ can be captured by CEFA. 
We start with the simple example for $v=a$.
\begin{example}[CEFA for $\indexof_a$] 
	Let $a \in \Sigma$. Then  $\CEFA_{\indexof_a} = (\{(q_0, q_1, q_2)\}, \Sigma, (r_1,r_2), \delta_{\indexof_a}, \{q_0\}, \{q_2\})$, where $\delta_{\indexof_a}$ comprises the tuples
	\begin{itemize}
		\item $(q_0, b, q_0, \eta)$ such that $b \in \Sigma$, $\eta(r_1)=1$, $\eta(r_2)=1$,
		\item $(q_0, b, q_1, \eta)$ such that $b \in \Sigma$, $\eta(r_1)=0$, $\eta(r_2) = 1$,
		\item $(q_0, a, q_2, \eta)$ such that $\eta(r_1)=0$, $\eta(r_2) = 0$,
		\item $(q_1, b, q_1, \eta)$ such that $b \in \Sigma \setminus \{a\}$, $\eta(r_1)=0$, $\eta(r_2)=1$,
		\item $(q_1, a, q_2, \eta)$ such that $\eta(r_1)=0$, $\eta(r_2)=0$,
		\item $(q_2, b, q_2, \eta)$ such that $b \in \Sigma$, $\eta(r_1)=0$, $\eta(r_2)=0$.
	\end{itemize}
	Intuitively, $r_1$ corresponds to the starting position $i$ of $\indexof_a(x, i)$, $r_2$ corresponds to the output of $\indexof_a(x, i)$, $q_0$ specifies that the current position is before $i$, $q_1$ specifies that the current position is after $i$, while $a$ has not occurred yet, and $q_2$ specifies that $a$ has occurred after $i$. 
\end{example}

Technically, for any NFA $\NFA$ and constant string $v$, we can construct a CEFA accepting $\{(w, (n, \indexof_v(w, n)))\mid w\in \Lang(\NFA), n \le \indexof_v(w, n) < |w| \}$. 
For this purpose, we need a concept of window profiles of  string positions w.r.t. $v$, which are elements of $\{\bot, \top\}^{n-1}$. The window profiles facilitate recognising the first occurrence of $v$ in the input string.  Intuitively, given a string $u$, the window profile of a position $i$ in $u$ w.r.t. $v$ encodes the matchings of prefixes of $v$ to the suffixes of $u[0,i]$ (see \cite{CCH+18} for the details). For $\pi = \pi_1 \cdots \pi_{n-1} \in \{\bot, \top\}^{n-1}$ and $b \in \Sigma$, we use $\uwp(\vec{\pi}, b)$ to represent the window profile updated from $\pi$ after reading the letter $b$, specifically, $\uwp(\vec{\pi}, b) = \vec{\pi'}$ such that  
\begin{itemize}
\item $\pi'_1 = \top$ iff $b = a_1$, 
\item for each $i \in [n-2]$, $\pi'_{i+1} = \top$ iff $\pi_{i} = \top$ and $b = a_{i+1}$. 
\end{itemize}
Let $WP_v$ denote the set of window profiles of string positions w.r.t. $v$. From the result in \cite{CCH+18}, we know that $|WP_v| \le |v|$. 

Suppose $v = a_1 \cdots a_n$ with $n \ge 2$. 
Then $\indexof_v$ is captured by the CEFA $\CEFA_{\indexof_v}=(Q, \Sigma, R, \delta, I, F)$, such that 
\begin{itemize}
\item $Q = \{q_0, q_1\} \cup WP_v \cup WP_v \times [n]$, 
\item $R=(r_1, r_2)$ (where $r_1,r_2$ represent the input and output positions of $\indexof_v$ respectively), 
\item $I=\{q_0\}$, 
\item $F=\{q_1\}$, and 
\item $\delta$ comprises 
\begin{itemize}
\item the tuples $(q_0, a, q_0, \eta)$ such that $a \in \Sigma$, $\eta(r_1)=1$, and $\eta(r_2) = 1$,
\item the tuples $(q_0, a, \vec{\pi}, \eta)$ such that $a \in \Sigma$, $\vec{\pi} = \theta \bot^{n-2}$ where $\theta  = \top$ iff $a = a_1$, $\eta(r_1) = 0$, and $\eta(r_2)= 0$ (recall that the first position of a string is $0$),
\item the tuples  $(\vec{\pi}, a, \uwp(\vec{\pi}, a), \eta)$ such that $\vec{\pi} \in WP_u$, $a \in \Sigma$, $\pi_{n-1} = \bot$ or $a \neq a_{n}$, $\eta(r_1) = 0$, and $\eta(r_2)= 1$,
\item the tuples $(\vec{\pi}, a, (\uwp(\vec{\pi}, a), 1), \eta)$ such that $\vec{\pi} \in WP_u$, $a = a_1$, $\pi_{n-1} = \bot$ or $a \neq a_{n}$, $\eta(r_1) = 0$, and $\eta(r_2)= 1$,
\item the tuples $((\vec{\pi}, i),  a, (\uwp(\vec{\pi}, a), i+1), \eta)$ such that $\vec{\pi} \in WP_u$, $i \in [n-2]$, $a = a_{i+1}$, $\pi_{n-1} = \bot$ or $a \neq a_{n}$, $\eta(r_1) = 0$, and $\eta(r_2)= 0$,
\item the tuples $((\vec{\pi}, n-1),  a, q_1, \eta)$ such that $\vec{\pi} \in WP_u$, $a = a_{n}$, $\eta(r_1) =0$, and $\eta(r_2)= 0$,
\item the tuples  $(q_1, a, q_1, \eta)$ such that $a \in \Sigma$, $\eta(r_1) = 0$, and $\eta(r_2)= 0$.
\end{itemize}
\end{itemize}


\section{Proof of Proposition~\ref{prop:pre-image}}\label{app:pre-image}

\noindent {\bf Proposition~\ref{prop:pre-image}}.
\emph{Let $L$ be a CERL defined by a CEFA $\CEFA = (Q, \Sigma, R, \delta, I, F)$. Then for each string function $f$ ranging over $\concat$, $\replaceall_{e,u}$, $\reverse$, FFTs $\NFT$, and $\substring$, $f^{-1}_R(L)$ is CERR-definable. In addition,
\begin{itemize}
\item a CEFA representation of $\concat^{-1}_R(L)$ can be computed in time $\bigO(|\CEFA|^2)$, 
\item a CEFA representation of $\reverse^{-1}_R(L)$ (resp. $\substring^{-1}_R(L)$) can be computed in time $\bigO(|\CEFA|)$,
\item a CEFA representation of  $(\Tran(\NFT))^{-1}_R(L)$ can be computed in time polynomial in $|\CEFA|$ and exponential in $|\NFT|$,
\item a CEFA representation of  $(\replaceall_{e,u})^{-1}_R(L)$ can be computed in time polynomial in $|\CEFA|$ and exponential in $|e|$ and $|u|$.
\end{itemize}
}

\begin{proof}
	Let $\CEFA=(Q, \Sigma, R, \delta, I, F)$ be a CEFA with $R= (r_1, \cdots, r_k)$. We show how to construct a CEFA representation of $f^{-1}_R(L)$ for each function $f$ in {\slint}.
	
	\paragraph*{$\concat^{-1}_R(L)$.}
	A CEFA representation of $\concat^{-1}_R(L)$ is given by $((\CEFA_{I, q}, \NFA_{q, F})_{q \in Q}, \vec{t})$, where 
	\begin{itemize}
		\item $\CEFA_{I, q}=(Q, \Sigma, R^{(1)}, \delta^{(1)}, I, \{q\})$ and  $\CEFA_{q, F}=(Q, \Sigma, R^{(2)}, \delta^{(2)}, \{q\}, F)$ such that 
		\begin{itemize}
			\item $R^{(1)} = (r^{(1)}_1, \cdots, r^{(1)}_k)$, $R^{(2)} = (r^{(2)}_1, \cdots, r^{(2)}_k)$, 
			\item $\delta^{(1)}$ comprises the tuples $(q, a, q', \eta')$ satisfying that there exists $\eta$ such that $(q, a, q', \eta) \in \delta$ and for each $j \in [k]$, and $\eta'(r^{(1)}_j)=\eta(r_j)$,  similarly for $\delta^{(2)}$,
		\end{itemize}
		\item and $\vec{t} = (r^{(1)}_1 + r^{(2)}_1, \cdots, r^{(1)}_k + r^{(2)}_k)$.
	\end{itemize}
	Note that the size of $((\CEFA_{I, q}, \NFA_{q, F})_{q \in Q}, \vec{t})$ is $\bigO(|\CEFA|^2)$.
	%
	%
	\paragraph*{$\reverse^{-1}_R(L)$.} 
	A CEFA representation of $\reverse^{-1}_R(L)$ is given by $(\CEFA^{(r)}, \vec{t})$, where 
	\begin{itemize}
		\item $\CEFA^{(r)}=(Q, \Sigma, R^{(1)}, \delta', F, I)$ such that 
		\begin{itemize}
			\item $R^{(1)}=(r^{(1)}_1,\cdots,r^{(1)}_k)$, and 
			\item $\delta'$ comprises the tuples $(q', a, q, \eta')$ satisfying that there exists $\eta$ such that $(q, a, q', \eta) \in \delta$, and $\eta'(r^{(1)}_i) = \eta(r_i)$ for each $i \in [k]$,
		\end{itemize}
		\item and $\vec{t}=(r^{(1)}_1, \cdots, r^{(1)}_k)$. 
	\end{itemize}
	Note that $\Lang(\CEFA^{(r)}) = \{(w^{(r)}, \vec{n}) \mid (w, \vec{n}) \in \Lang(\CEFA)\}$, and the size of $(\CEFA^{(r)}, \vec{t})$ is $\bigO(|\CEFA|)$.
	
	\paragraph*{$\substring^{-1}_R(L)$.}
	A CEFA representation of $\substring^{-1}_R(L)$ is given by $(\cB, \vec{t})$, where 
	\begin{itemize}
		\item $\cB = (Q', \Sigma, R', \delta', I', F')$ such that 
		\begin{itemize}
			\item $Q' = Q \times \{p_0, p_1, p_2\}$, (intuitively, $p_0$, $p_1$, and $p_2$ denote that the current position is before the starting position, between the starting position and ending position, and after the ending position respectively)
			\item $R' = \left(r'_{1,1}, r'_{1,2}, r^{(1)}_1,\cdots, r^{(1)}_k \right)$, (intuitively, $r'_{1,1}$ denotes the starting position, and $r'_{1,2}$ denotes the length of the substring)
			\item $I'=I \times \{p_0\}$, $F'=F' \times \{p_2\} \cup (I \cap F) \times \{p_0\}$,
			\item and $\delta'$ comprises 
			\begin{itemize}
				\item the tuples $((q, p_0), a, (q, p_0), \eta')$ such that $q \in I$, $a \in \Sigma$, and $\eta'$ satisfies that $\eta'(r'_{1,1})= 1$, and $\eta'(r'_{1,2}) = 0$, and $\eta'(r^{(1)}_j)=0$ for each $j \in [k]$, 
				\item the tuples $((q, p_0), a, (q', p_1), \eta')$ such that $q \in I$ and there exists $\eta$ satisfying that $(q, a, q', \eta) \in \delta$, moreover, $\eta'(r'_{1,1})=0$ (recall that the positions of strings start at $0$), $\eta'(r'_{1,2}) = 1$, and $\eta'(r^{(1)}_j)=\eta(r_j)$ for each $j \in [k]$,
				\item the tuples $((q, p_0), a, (q', p_2), \eta')$ such that $q \in I$ and there exists $\eta$ satisfying that $(q, a, q', \eta) \in \delta$, moreover, $q' \in F$, and $\eta'(r'_{1,1})=0$ (recall that the positions of strings start at $0$), $\eta'(r'_{1,2}) = 1$, and $\eta'(r^{(1)}_j)=\eta(r_j)$ for each $j \in [k]$,  
				\item the tuples $((q, p_1), a, (q', p_1), \eta')$ such that there exists $\eta$ satisfying that $(q, a, q', \eta) \in \delta$, $\eta'(r'_{1,1}) = 0$, and $\eta'(r'_{1,2}) = 1$, and $\eta'(r^{(1)}_j)=\eta(r_j)$ for each $j \in [k]$,
				\item the tuples $((q, p_1), a, (q', p_2), \eta')$ such that $q' \in F$, and there exists $\eta$ satisfying that $(q, a, q', \eta) \in \delta$, moreover, $\eta'(r'_{1,1}) = 0$, $\eta'(r'_{1,2}) = 1$, and $\eta'(r^{(1)}_j)=\eta(r_j)$ for each $j \in [k]$,
				%
				%
				\item the tuples $((q, p_2), a, (q, p_2), \eta')$ such that $q \in F$, $\eta'(r'_{1,1}) = 0$, and $\eta'(r'_{1,2}) = 0$, and $\eta'(r^{(1)}_j)=0$ for each $j \in [k]$,
			\end{itemize}
		\end{itemize}
		\item $\vec{t}=(r^{(1)}_1, \cdots, r^{(1)}_k)$.
	\end{itemize}
	Note that the size of $(\cB, \vec{t})$ is $\bigO(|\CEFA|)$.
	%
	%
	\paragraph*{$(\Tran(\NFT))^{-1}_R(L)$.}
	Suppose $\NFT = (Q', \Sigma, \delta', I', F')$. Then a CEFA representation of $(\Tran(\NFT))^{-1}_R(L)$ is given by 
	$(\cB, \vec{t})$, where 
	\begin{itemize}
		\item $\cB$ simulates the run of $\NFT$ on the input string, meanwhile, it simulates the run of $\CEFA$ on the output string of $\NFT$, formally, $\cB= (Q' \times Q, \Sigma, R^{(1)}, \delta'', I' \times I, F' \times F)$ such that 
		\begin{itemize}
			\item $R^{(1)}  = (r^{(1)}_1, \cdots, r^{(1)}_k)$, and
			\item $\delta''$ comprises the tuples $((q'_1, q_1), a, (q'_2, q_2), \eta')$ satisfying one of the following conditions,
			\begin{itemize}
				\item there exist $u = a_1 \cdots a_n \in \Sigma^+$ and a transition sequence $p_0 \xrightarrow[\delta]{a_1, \eta_1} p_2 \cdots p_{n-1} \xrightarrow[\delta]{a_n, \eta_n} p_{n}$ in $\CEFA$ such that $(q'_1, a, q'_2, u) \in \delta'$, $p_0 = q_1$, $p_{n}= q_2$, and for each $j \in [k]$,  $\eta'(r^{(1)}_j) = \eta_1(r_j) + \cdots + \eta_n(r_j)$,
				\item $(q'_1, a, q'_2, \varepsilon) \in \delta'$, $q_1 = q_2$, and $\eta'(r^{(1)}_j) =0$ for each $j \in [k]$,
			\end{itemize}
		\end{itemize}
		\item $\vec{t}=(r^{(1)}_1, \cdots, r^{(1)}_k)$.
	\end{itemize}
	Note that the number of transitions of $\cB$ can be exponential in the worst case, since it summarises the updates of cost registers of $\CEFA$ on the output strings of the transitions of $\NFT$. More precisely,  let
	\begin{itemize}
		\item $\ell$ be the maximum length of the output strings of transitions of $\NFT$, 
		\item $N$ be the maximum number of transitions between a given pair of states of $\CEFA$, and
		\item  $C$ be the maximum absolute value of the integer constants occurring in $\CEFA$,
	\end{itemize}
	then $|\delta''|$, the cardinality of $\delta''$, is bounded by $|\delta'| \times |Q|^2 \times N^\ell $, and the integer constants occurring in each transition of $\delta''$ are bounded by $\ell C$. Therefore, 
	the size of $(\cB, \vec{t})$ is 
	\[
	\bigO(|\delta'| \times |Q|^2 \times N^\ell \times k \log_2 (\ell C)).
	\] 
	Since $|\delta'|, \ell \le |\NFT|$, $|Q|, N, k \le |\CEFA|$, and $C \le 2^{|\CEFA|}$, we deduce that the size of $(\cB, \vec{t})$ is 
	$
	\bigO( |\NFT| \times  |\CEFA|^2 \times |\CEFA|^{|\NFT|} \times |\CEFA|^2 \log_2(|\NFT|))= |\CEFA|^{\bigO(|\NFT|)} |\NFT| \log_2(|\NFT|).$
	%
	
	\paragraph*{$(\replaceall_{e,u})^{-1}_R(L)$.}
	From the result in \cite{CCH+18}, we know that  a NFT $\NFT_{e,u}=(Q', \Sigma, \delta', I', F')$ can be constructed to capture $\replaceall_{e,u}$.  Moreover, 
	\begin{itemize}
		\item $|Q'|$, as well as $|\delta'|$, is $2^{\bigO(|e|)}$,
		\item $\ell$, the maximum length of the output strings of transitions of $\NFT_{e,u}$, is $|u|$.
	\end{itemize}
	Then a CEFA representation of $(\replaceall_{e,u})^{-1}_R(L)$ can be constructed as that of $(\Tran(\NFT_{e,u}))^{-1}_R(L)$.
	Let $N$ denote the maximum number of transitions between a given pair of states of $\CEFA$, and $C$ be the maximum absolute value of the integer constants occurring in $\CEFA$, which is bounded by $2^{|\CEFA|}$. Then the CEFA representation of $(\replaceall_{e,u})^{-1}_R(L)$ is of size 
	\[
	\bigO(|\delta'| \times |Q|^2 \times N^\ell \times k \log_2 (\ell C)) = 2^{\bigO(|e|)} |\CEFA|^2 |\CEFA|^{|u|} |\CEFA|^2 \log_2 |u|=2^{\bigO(|e|)} |\CEFA|^{\bigO(|u|)}.
	\]
	according to the aforementioned discussion for NFTs.
\end{proof}

\section{Proof of Proposition~\ref{prop:la-sat-cefa-inter}}\label{app:sat-cefa}

\noindent{\bf Proposition~\ref{prop:la-sat-cefa-inter}}.
\emph{The {\lasat} problem is $\pspace$-complete.}

\begin{proof}
	The lower bound follows from the {\pspace}-hardness of the intersection problem of NFAs. 
	
	For the upper bound, let $\{ \CEFA_i^{j} \}_{i\in I,j\in J_i}$ be a family of CEFAs  each of which carries a vector of registers $R_i^j$ and  $\phi$ be a quantifier-free LIA formula such that  $ R_i^{j} $ are pairwise disjoint and the variables of $\phi$ are from $R':=\bigcup_{i,j} R_i^j$. 
	
	First, we observe that we can focus on \emph{monotonic CEFAs} where the cost registers are monotone in the sense that their values are non-decreasing during the course of execution. In other words, they can only be updated with natural number (as opposed to general integer) constants. This observation is justified by the following reduction.
	
	For each register $r \in R^i_j$, we introduce two registers $r^+, r^-$. Let $(R^i_j)^{\pm}$ denote the vector of registers by replacing each $r \in R^i_j$ with $(r^+, r^-)$. Intuitively,  for each $r \in R^i_j$, the updates of $r$ in $\CEFA_i^{j} $ are split into non-negative ones and negative ones, with the former stored in $r^+$ and the latter in $r^-$. Suppose $(R')^{\pm} = \bigcup_{i,j} (R_i^j)^{\pm}$. Then we construct monotonic CEFAs $(\cB_i^{j})_{i \in I, j \in J_i}$ and an LIA formula $\phi^\pm$ such that
	\begin{quote}
		there are an assignment function $\theta: R' \rightarrow \Int$ and strings $(w_i)_{i \in I}$ such that  $\phi[\theta(R' )/R']$ holds and $(w_i, \theta(R_i^j)) \in \Lang(\CEFA_{i}^j)$ for every $i \in I$ and $j \in J_i$ 
		\begin{center} if and only if \end{center}
		there are an assignment function $\theta^\pm: (R')^\pm \rightarrow \Nat$ and strings $(w_i)_{i \in I}$ such that  $\phi^\pm[\theta^\pm((R')^\pm)/(R')^\pm]$ holds and $(w_i, \theta^\pm((R_i^j)^\pm)) \in \Lang(\cB_{i}^j)$ for every $i \in I$ and $j \in J_i$.
	\end{quote}
	For $i \in I$ and $j \in J_i$, the CEFA $\cB_{i}^j$ is obtained from $\CEFA_{i}^j$ by replacing each transition $(q, a, q', \eta)$ in $\CEFA_i^j$ by the transition $(q, a, q', \eta')$ such that for each $r \in R_j^j$, 
	\[
	\eta'(r^+) = \left\{ \begin{array}{l  l}\eta(r), & \mbox{ if } \eta(r) \ge 0 \\ 0 & \mbox{ otherwise} \end{array}\right.,  \eta'(r^-) = \left\{ \begin{array}{l  l} 0, & \mbox{ if } \eta(r) \ge 0 \\ -\eta(r) & \mbox{ otherwise} \end{array}\right..
	\]
	In addition, $\phi^\pm$ is obtained from $\phi$ by replacing each $r \in R'$ with $r^+-r^-$.
	
	\smallskip
	
	It remains to prove the {\lasat} problem for monotonic CEFAs is in {\pspace}, namely,
	\begin{quote}
		given a family of \emph{monotonic} CEFAs $\{ \CEFA_i^{j} \}_{i\in I,j\in J_i}$ each of which carries a vector of registers $R_i^j$ and a quantifier-free LIA formula $\phi$ such that  $ R_i^{j} $ are pairwise disjoint,  and the variables of $\phi$ are from $R'=\bigcup_{i,j} R_i^j$, deciding whether  there are an assignment function $\theta: R' \rightarrow \Nat$ and strings $(w_i)_{i \in I}$ such that  $\phi[\theta(R' )/R']$ holds and $(w_i, \theta(R_i^j)) \in \Lang(\CEFA_{i}^j)$ for every $i \in I$ and $j \in J_i$ is in {\pspace}.
	\end{quote}

	We use Proposition 16 in \cite{LB16} to show the result. Proposition 16 in \cite{LB16} mainly considered monotonic counter machines, which can be seen as monotonic CEFAs where each transition contains no alphabet symbol, and $\eta(r) \in \{0,1\}$ for the update function $\eta$ therein.
	
	For each $i\in I$ and $j\in J_i$, let $(\CEFA')_i^j$ be the monotonic counter machine obtained from $\CEFA_i^{j}$ by the following two-step procedure:
	\begin{enumerate}
		\item {[Remove the alphabet symbols]}: Remove alphabet symbols $a$ in each transition $(q, a, q', \eta)$ of $\CEFA_i^{j}$.
		\item {[From binary encoding to unary encoding]}: Replace each transition $(q, q', \eta)$ such that $\ell = \max_{r \in R_i^j} \eta(r) > 1$ with a sequence of transitions $(q, p_1,\eta'_1), \cdots, (p_{\ell-1}, q', \eta'_\ell)$, where $p_1, \cdots,p_{\ell-1}$ are the freshly introduced states, moreover, $\eta'_j(r) = 1$ if $\eta(r) \ge j$, and $\eta'_j(r)=0$ otherwise. 
	\end{enumerate}
	
	According to Proposition 16 in \cite{LB16}, we have the following property. 
	\begin{quote}
		Given a family of monotonic counter machines $\{ \cC_i \}_{i\in I}$ each of which carries a vector of counters $R_i$ and a quantifier-free LIA formula $\phi$ such that $ R_i$ are pairwise disjoint,  and the variables of $\phi$ are from $R'=\bigcup_{i} R_i$. If there is an assignment function $\theta: R' \rightarrow \Nat$ such that $\phi[\theta(R' )/R']$ holds and $\theta(R_i)$ is a reachable valuation of counters in $\cC_i$ for every $i \in I$, then there are desired $\theta$ such that for each $i \in I$ and $r \in R_i$, $\theta(r)$ is at most polynomial in the number of states in $\cC_i $, exponential in $|R_i|$, and exponential in $|\phi|$.
	\end{quote}
	For each $i \in I$, let $\cC_i$ be the product of monotonic counter machines $(\CEFA')_i^j$ for $j \in J_i$. 
	From the fact that the number of states of $(\CEFA')_i^j$ is at most the product of the number of transitions of $\CEFA_i^j$ and $B_{\CEFA_i^j}$ (where $B_{\CEFA_i^j}$ denotes the maximum natural number constants $\eta(r)$ in $\CEFA_i^j$), we deduce the following,
	\begin{quote}
		if there are an assignment function $\theta: R' \rightarrow \Nat$ and strings $(w_i)_{i \in I}$ such that  $\phi[\theta(R' )/R']$ holds and $(w_i, \theta(R_i^j)) \in \Lang(\CEFA_i^j)$ for every $i \in I$ and $j \in J_i$, then there are desired $\theta$ and $(w_i)_{i \in I}$ such that for each $i \in I$ and $r \in \bigcup_{j \in J_i} R^j_i$, $\theta(r)$ is at most polynomial in the product of the number of transitions in $\CEFA_i^j$ and $B_{\CEFA_i^j}$ for $j \in J_i$, exponential in $\left|\bigcup_{j \in J_i} R^j_i \right|$, and exponential in $|\phi|$.
	\end{quote}
	
	Since the values of all the registers in $\CEFA_i^j$ for $i \in I$ and $ j \in J_i$ can be assumed to be at most exponential, and thus their binary encodings can be stored in polynomial space, one can nondeterministically guess the strings $(w_i)_{i \in I}$, and for each $i \in I$ and $j \in J_i$, simulate the runs of CEFAs $\CEFA_i^j$ on $w_i$, and finally evaluate $\phi$ with the register values after all $\CEFA_{i}^{j}$ accept, in polynomial space. From Savitch's theorem \cite{complexity-book}, we conclude that the {\lasat} problem for monotonic CEFAs is in {\pspace}. This concludes the proof of the proposition.
\end{proof}

\section{Implementation} \label{appendix:impl}
%

\begin{algorithm}[htbp]
	\SetKw{Continue}{continue}
	\small
	\KwIn{$active$: set of CEFA constraints,  $arith$: arithmetic constraints,
		$\mathit{funApps}$: acyclic set of assignment statements. }
	\KwResult{$\mathit{sat}$ if the input constraints are satisfiable, and $\mathit{unsat}$ otherwise.\newline
	}
	\For{each partition $(\mathcal{I}_l)_{l \in [5]}$ of the set of $\indexof_v(x, i)$ in $\mathit{arith}$ and \newline
		\hspace*{4mm} each partition $(\mathcal{J}_l)_{l \in [3]}$ of the set of $\substring(x, i, j)$ in $\mathit{funApps}$ 
		\tcc{the partitions refer to (1)-(5) for $\indexof_v(x, i)$ and (1)-(3) for $\substring(x, i, j)$ in Step II of Section~\ref{sec:dc}}
	}
	{
		\tcc{Case splits for semantics of $\indexof$ and $\substring$}
		$(\mathit{active}, \mathit{arith}, \mathit{funApps}) = \mathit{indexofCaseSplit}(\mathit{active}, \mathit{arith}, \mathit{funApps}, (\mathcal{I}_l)_{l \in [5]})$\; 
		$(\mathit{active}, \mathit{arith}, \mathit{funApps})= \mathit{substringCaseSplit}(\mathit{active}, \mathit{arith}, \mathit{funApps}, (\mathcal{J}_l)_{l \in [3]})$\; 
		\For{each $\length(x)$ occurring in $\mathit{arith}$}
		{
			choose a fresh integer variable $i$\;
			$\mathit{active} \leftarrow \mathit{active} \cup \{x \in \CEFA_{\rm len}[i/r_1]\}$; $\mathit{arith}\leftarrow \mathit{arith}[i/\length(x)]$;
		}
		\For{each $\indexof_v(x,i)$ occurring in $\mathit{arith}$}
		{
			choose fresh integer variables $i_1,i_2$\;
			$\mathit{active} \leftarrow \mathit{active} \cup \{x \in \CEFA_{\indexof_v}[i_1/r_1,i_2/r_2]\}$; $\mathit{arith}\leftarrow \mathit{arith}[i_2/\indexof_v(x,i)] \wedge i=i_1$;
		}
		\If{$\mathit{BackDfsExp}(\mathit{active}, \emptyset, \mathit{arith}, \mathit{funApps})$}
		{
			\Return{$sat$};}
	}
	\Return $\mathit{unsat}$; 		
	\caption{Function $\mathit{checkSat}$
		for Step II-III} \label{alg:checksat} 
\end{algorithm}

OSTRICH+ performs a depth-first exploration of the search tree resulting from repeatedly
splitting the disjunctions (or unions) in the cost-enriched recognisable pre-images of CERLs under string functions, as well as the case splits in the semantics of $\indexof$ and $\substring$. The pseudo-code of Step II-III of the decision procedure 
is given by  the function $\mathit{checkSat}$ in Algorithm~\ref{alg:checksat}, which calls two functions $\mathit{indexofCaseSplit}$ in Algorithm~\ref{alg:indexof} and $\mathit{substringCaseSplit}$ in Algorithm~\ref{alg:substring} for the case splits in the semantics of $\indexof_v$ and $\substring$ respectively.
%
%
Moreover,  $\mathit{checkSat}$ calls a recursive function  $\mathit{BackDfsExp}$ in Algorithm~\ref{alg:dfs} for the depth-first exploration (Step IV of the decision procedure), which in turn calls a function $\mathit{CheckCefaLIASat}$ to solve the {\lasat} problem (Step V). Note that Step I of the decision procedure is handled by the DPLL(T) procedure in Princess and is omitted here.





\paragraph*{Optimisations for solving the {\lasat} problem.} From Proposition~\ref{prop:la-sat-cefa-inter}, a natural approach to solve the {\lasat} problem is to compute an existential LIA formula defining the Parikh image of products of CEFAs, and then use off-the-shelf SMT solvers (e.g. CVC4 or Z3) to decide the satisfiability of the existential LIA formula. 
However, our preliminary experiments show that this approach suffers from a scalability issue, in particular, 
the state-space explosion when computing products of CEFAs.  
In the implementation of the function $\mathit{CheckCefaLIASat}$ in Algorithm~\ref{alg:dfs},  we 
opt to utilise the symbolic model checker nuXmv \cite{nuxmv} to mitigate the state-space explosion during the computation of products of CEFAs. The nuXmv tool is a well-known symbolic model checker that is capable of analysing both finite and infinite state systems. Our technique is to encode {\lasat} as an instance of the model checking problem, which can  be solved by 
nuXmv. Since  {\lasat} is a problem for quantifier-free LIA formulas and CEFAs that contain integer variables, the {\lasat} problem actually corresponds to the problem of model checking \emph{infinite state systems}. 
%
%

\vspace{-5mm}
\begin{algorithm}[htbp]
	\small
	\KwIn{$active$: set of CEFA constraints,  $arith$: arithmetic constraint,
		$\mathit{funApps}$: acyclic set of assignment statements, and $(\mathcal{I}_l)_{l \in [5]}$: subsets of $\indexof_v(x,i)$ string terms}
	\KwResult{$(active, arith, \mathit{funApps})$\newline}
	
	\For{each $\indexof_v(x, i) \in \mathcal{I}_1$}
	{
		$arith \leftarrow arith[\indexof_v(x, 0)/\indexof_v(x,i)] \wedge i < 0$\;
	}
	\For{each $\indexof_v(x, i) \in \mathcal{I}_2$}
	{
		$active \leftarrow active \cup \{x \in \NFA_{\overline{\Sigma^* v \Sigma^*}}\}$\;
		$arith \leftarrow arith[-1/\indexof_v(x,i)] \wedge i < 0$\;
	}
	\For{each $\indexof_v(x, i) \in \mathcal{I}_3$}
	{
		$arith \leftarrow arith[-1/\indexof_v(x,i)] \wedge i \ge \length(x)$\;
	}
	\For{each $\indexof_v(x, i) \in \mathcal{I}_4$}
	{
		$arith \leftarrow arith[-1/\indexof_v(x,i)] \wedge i \ge 0 \wedge i < \length(x)$\;
	}
	\For{each $\indexof_v(x, i) \in \mathcal{I}_5$}
	{
		choose fresh variables $y$ and $j$\;
		$active \leftarrow active \cup \{y \in \NFA_{\overline{\Sigma^* v \Sigma^*}}\}$\;
		$arith \leftarrow arith[-1/\indexof_v(x,i)] \wedge i \ge 0 \wedge i < \length(x) \wedge j = \length(x)-i$\;
		$\mathit{funApps} \leftarrow \mathit{funApps} \cup \{y:=\substring(x, i, j)\}$\;
	}		
	\caption{$\mathit{indexofCaseSplit}$ for case splits in the semantics of $\indexof_v$}\label{alg:indexof}
\end{algorithm}

\begin{algorithm}[htbp]
	\small
	\KwIn{$active$: set of CEFA constraints,  $arith$: arithmetic constraint,
		$\mathit{funApps}$: acyclic set of assignment statements, and $(\mathcal{I}_l)_{l \in [5]}$: subsets of $\indexof_v(x,i)$ string terms}
	\KwResult{$(active, arith, \mathit{funApps})$\newline}
	
	\For{each $y:=\substring(x, i, j) \in \mathcal{J}_1$}
	{
		$arith \leftarrow arith \wedge i \ge 0 \wedge i+j \le \length(x)$;
	}
	\For{each $y:=\substring(x, i, j) \in \mathcal{J}_2$}
	{
		choose a fresh integer variable $i'$\;
		$arith \leftarrow arith \wedge i \ge 0 \wedge i \le \length(x) \wedge i+j > \length(x) \wedge i' = \length(x)-i$\;
		$\mathit{funApps} \leftarrow \mathit{funApps}[y:=\substring(x, i, i')/y:=\substring(x, i, j)]$\;
	}
	\For{each $y:=\substring(x, i, j) \in \mathcal{J}_3$}
	{
		$arith \leftarrow arith \wedge i < 0$\;
		$active \leftarrow active \cup \{y \in \NFA_\varepsilon\}$\;
		$\mathit{funApps} \leftarrow \mathit{funApps} \setminus \{y:=\substring(x, i, j)\}$\;		 
	}
	\caption{$\mathit{substringCaseSplit}$  for case splits in the semantics of $\substring$}\label{alg:substring}
\end{algorithm}
\vspace*{-1cm}

\begin{algorithm}[htbp]
\SetKw{Continue}{continue}
  \small
  \KwIn{$\mathit{active}, \mathit{passive}$: sets of CEFA constraints,  $\mathit{arith}$: arithmetic constraints,
    $\mathit{funApps}$: acyclic set of assignment statements. }
  \KwResult{$\mathit{sat}$ if the input constraints are satisfiable, and $\mathit{unsat}$ otherwise.\newline
   }
%
    \eIf{$\mathit{active} = \emptyset$}{
      \tcc{Check whether the LIA constraint $\mathit{arith}$ is satisfiable with respect to the CEFA constraints in $\mathit{passive}$ (i.e. Step V).}
      \Return{$\mathit{CheckCefaLIASat}(\mathit{passive}, \mathit{arith})$;} 
    }{
   	choose a CEFA constraint $x \in \CEFA$ in $active$ with $R(\CEFA)=(r_1,\cdots,r_k)$\;
	\eIf{there is an assignment~$x := f(y_1, \vec{i_1}, \ldots, y_l,\vec{i_l})$ defining $x$ in $\mathit{funApps}$ with \newline
	\hspace*{4mm} $\vec{i_j}=(i_{j,1},\cdots, i_{j, k_j})$ for $j \in [l]$
	}
	{
		compute $f^{-1}_{R(\CEFA)}(\Lang(\CEFA)) = \left((\CEFA^{(1)}_{j}, \cdots, \CEFA^{(l)}_{j})_{j \in [n]}, \vec{t}\right)$ where \newline 
		 $R\left(\CEFA^{(j')}_{j}\right)=\left((r')^{(j', 1)}, \cdots,(r')^{(j', k_{j'})}, r^{(j')}_1,\cdots, r^{(j')}_k \right)$ for $j \in [n]$ and $j' \in [l]$\;
	        $\mathit{active} \leftarrow \mathit{active} \setminus \{x \in \CEFA\}$; $\mathit{passive} \leftarrow \mathit{passive} \cup \{x \in \CEFA\}$\;    
	        \For{$j \leftarrow 1$ \KwTo $n$}{
        		$\mathit{active} \leftarrow \mathit{active} \cup \{y_1 \in \CEFA^{(1)}_{j}, \ldots, y_l \in \CEFA^{(l)}_{j}\} $\;
        		$\mathit{arith} \leftarrow \mathit{arith} \wedge \bigwedge_{j' \in [l], j'' \in [k_{j'}]} i_{j',j''} = (r')^{(j', j'')} \wedge \bigwedge_{j' \in [k]} r_{j'} = t_{j'}$\;
        		\eIf{$\mathit{active} \cup \mathit{passive}$ is inconsistent}{\Continue \tcc*{backtrack}}
		{
		          \Switch{$\mathit{BackDfsExp}(\mathit{active}, \mathit{passive}, \mathit{arith}, \mathit{funApps})$}{
				\lCase{$sat$}
					{\Return{$\mathit{sat}$}}
				\Case{$\mathit{unsat}$}
					{\Continue \tcc*{backtrack}}
          			}
        		}
	}
        	\Return{$\mathit{unsat}$}; 
      	}
      	{
        	\Return{$\mathit{BackDfsExp}(\mathit{active} \backslash \{x\in \CEFA\}, \mathit{passive} \cup \{x\in \CEFA\}, \mathit{arith}, \mathit{funApps})$;}
      	}
	} 
  \caption{Function~$\mathit{BackDfsExp}$ for Step IV (depth-first exploration)}\label{alg:dfs}
\end{algorithm}

\end{appendix}
}
{}

\end{document}